\documentclass[onecolumn,aps,amsmath,amssymb,superscriptaddress]{revtex4}

\usepackage{dcolumn}
\usepackage{bm}
\usepackage{amsmath}
\usepackage{amsfonts}
\usepackage{mathrsfs}  
\usepackage{graphics}
\usepackage[dvips]{graphicx}
\usepackage{times}
\usepackage{setspace}
\usepackage{color}   
\usepackage{verbatim}
\usepackage[raggedright]{titlesec}
\usepackage[normalem]{ulem}
\usepackage{framed}
\usepackage{enumerate}
\usepackage[breaklinks=true]{hyperref}
\usepackage{breakurl}

\usepackage{balance}

\usepackage{flushend}

\hypersetup{colorlinks=true,citecolor=blue, linkcolor=black,urlcolor=blue} 

\begin{document}

\title{Renormalizing individual performance metrics\\ for cultural heritage management of sports records}
\author{Alexander M. Petersen}
\affiliation{Management of Complex Systems Department, Ernest and Julio Gallo Management Program, School of Engineering, University of California, Merced, CA 95343 }
\author{Orion Penner}
\affiliation{Chair of Innovation and Intellectual Property Policy, College of Management of Technology, Ecole Polytechnique Federale de Lausanne, Lausanne, Switzerland.}

\begin{abstract}
Individual performance metrics are commonly used to compare players from different eras. However, such cross-era comparison is often biased due to significant changes in success factors underlying player achievement rates (e.g. performance enhancing drugs and modern training regimens). Such historical comparison  is more than fodder for casual discussion among sports fans, as it is also an issue of critical importance to the multi-billion dollar professional sport industry and the institutions (e.g. Hall of Fame) charged with preserving sports history and the legacy of outstanding players and achievements. To address this cultural heritage management issue,  we report an objective statistical method for renormalizing career achievement metrics, one that is particularly tailored  for  common seasonal  performance metrics, which are often aggregated  into summary career metrics -- despite the fact that many player careers span different eras. Remarkably, we find that the method applied to comprehensive Major League Baseball and National Basketball Association player data  preserves the overall functional form of the  distribution of career achievement, both at the season and career level. As such, subsequent   re-ranking of the top-50 all-time records in MLB and the NBA  using renormalized metrics indicates reordering at the local rank level, as opposed to bulk reordering by era. This local order refinement signals time-independent mechanisms underlying annual and  career achievement in professional sports, meaning that appropriately  renormalized achievement metrics can be used to compare players from eras with different season lengths, team strategies, rules -- and possibly even different sports. 
\end{abstract}

\date{\today}
\maketitle

\newcommand{\beginsupplement}{%
        \setcounter{section}{0} 
        \renewcommand{\thesection}{S\arabic{section}}%
        \setcounter{table}{0}
        \renewcommand{\thetable}{S\arabic{table}}%
        \setcounter{figure}{0}
        \renewcommand{\thefigure}{S\arabic{figure}}%
        \setcounter{equation}{0}
        \renewcommand{\theequation}{S\arabic{equation}}%
        \setcounter{page}{1}
     }

\footnotetext[1]{ Please send correspondence to: Alexander M. Petersen (apetersen3@ucmerced.edu) or Orion Penner (orion.penner@gmail.com)}
 
  \vspace{-0.2in}
\section*{Introduction}
\vspace{-0.2in}
Individual achievement in competitive endeavors -- such as professional sports \cite{petersen2008distribution,saavedra2010mutually,petersen2011methods,radicchi2011best,mukherjee2012identifying,mukherjee2014quantifying,yucesoy2016untangling}, academia \cite{radicchi2009diffusion,petersen2011quantitative,petersen2012persistence,petersen_citationinflation_2018}  and other  competitive arenas \cite{radicchi2012universality,schaigorodsky2014memory,liu2018hot,barabasi2018formula} -- depends on many factors. Importantly, some factors are time dependent whereas others are not. Time dependent factors can derive from overall policy  (rule changes) and biophysical shifts  (improved nutrition and training techniques), to competitive group-level determinants (e.g. talent dilution of players from league expansion, and shifts in the use of backup  players) and  individual-specific enhancements (performance enhancing drugs (PEDs) \cite{mitchell2007report,mazanov2010rethinking} and even cognitive enhancing drugs (CEDs) \cite{sahakian2007professor,maher2008poll,greely2008towards}). Accounting for era-specific factors in cross-era comparison (e.g.  ranking ) and decision-making (e.g. election of players to the Hall of Fame) is a  challenging problem for cultural heritage management in the present-day  multi-billion dollar  industry of professional sports. 

Here we analyze two prominent and longstanding sports leagues   -- Major League Baseball (MLB) and the National Basketball Association (NBA) --  which   feature  rich statistical game data, and consequently,  record-oriented fanbases \cite{ward1996baseball,simmons2010book}. Each sport has well-known measures of greatness, whether they are single-season benchmarks or career records, that implicitly assume that  long-term trends in player ability are negligible. However, this is frequently not the case, as a result of time-dependent endogenous and exogenous performance factors underlying competitive advantage and individual success in sport. Take for example the home run in baseball, for which the frequency (per-at-bat) has increased 5-fold from 1919 (the year that Babe Ruth popularized the achievement and took hold of the single-season record for another 42 years) to 2001 (when Barry Bonds hit 73 home runs, roughly 2.5 times as many as Ruth's record of 29 in 1919 \cite{petersen2008distribution}). Yet as this example illustrates, there is a measurement problem challenging the reverence of such {\it all-time} records, because it is implicitly  assumed that the underlying success rates are stationary (i.e. the average, standard deviation and higher-order moments of success rates are time-independent), which is likely not  the case -- especially when considering the entire history of a sport.

Indeed, this fundamental measurement problem is further compounded when considering  career metrics, which for many great athletes span multiple decades of play, and thus possibly span distinct eras defined by specific events (e.g. the 1969 lowering of the pitching mound in Major League Baseball which notably reduced the competitive advantage of pitchers, and the introduction  of the 3-point line  to the NBA in 1979). By way of example, consider again the comparison of Barry Bonds (career years 1986-2007) and Babe Ruth (1914-1935). Despite the fact that Barry Bonds is also the career-level home-run leader (762 home runs total; see Supplementary Material Appendix Table S1), one could  argue that since other contemporaneous sluggers during the `steroids era' (the primary era during which Bonds primarily payed)  were also hitting home-runs at relatively high rates, that these nominal achievements were relatively less outstanding  -- in a statistical sense -- compared to players from other eras when baseline home-run rates were lower. Thus, if the objective is to identify achievements that are outstanding relative to both contemporaneous peers in addition to all historical predecessors, then  standardized measures of achievement  that  account for the time-dependent performance factors are needed. 

In general, we argue that in order to compare human achievements from different time periods, success metrics should be {\it renormalized}  to a common index (also termed `detrended' or `deflated' in other domains \cite{petersen2011methods,petersen_citationinflation_2018,petersen2018mobility}), so that the time dependent factors do not bias statistical comparison.
Hence, we address this measurement problem by  leveraging  the annual distributions of individual player achievement derived from comprehensive  player data comprised of more than 21,000 individual  careers spanning the entire history of both MLB and the NBA through the late 2000s for which data is collected \cite{BaseballData,BasketballData}. More specifically, we  apply an intuitive statistical  method that neutralizes  time-dependent factors by renormalizing players' annual achievements  to an annual inter-temporal average measuring characteristic player {\it prowess} -- operationalized as  ability per in-game opportunity. In simple terms, this method corresponds to a simple rescaling of the achievement metric baseline. We show that this method succeeds in part due to the relatively stable functional form of the annual performance  distributions for the seven  performance metrics we analyzed: batter home runs (HR), batter hits (H), pitcher strikeouts (K) and pitcher wins (W) for  MLB; and points scored (Pts.), rebounds (Reb.) and assists (Ast.) for the NBA. As a result, the outputs of our renormalization method are self-consistent  achievement metrics that are more appropriate for comparing and evaluating the relative achievements of players from different historical eras. 

In order to make our statistical analysis accessible, we use the most natural measures for accomplishment -- the statistics that are listed in typical box-scores and  on every baseball and basketball card, so that the results are tangible to historians and casual fans interested in reviewing and discussing the ``all-time greats.''  Without loss of generality, our method can readily be applied to more sophisticated composite measures that are increasingly prevalent in sports analytics (e.g. `Win Shares' in baseball \cite{james2002win}).  However, other sophisticated measures that incorporate team-play data (e.g. Box Plus Minus for basketball) or context-specific play data (e.g. Wins Above Replacement for baseball) are less feasible  due to the difficulty in obtaining  the necessary game-play information, which that is typically not possible to reconstruct from crude newspaper boxscores, and thus limits the feasibility of performing  comprehensive historical analysis. 

Notwithstanding these limitations,  this study addresses two relevant  questions:

\begin{enumerate}
\item How to quantitatively account for economic, technological, and social factors that influence the rate of achievement in competitive professions.

\item How to objectively compare individual career metrics for players from distinct historical eras. By way of example, this method could facilitate both standard and
retroactive induction of athletes into  Halls of Fame. This is particularly relevant  given the  `inflation' in the home run rate observed in Major League Baseball during the `steroids era' \cite{petersen2008distribution,mitchell2007report}, and the overarching  challenges of accounting for PEDs and other paradigm shifts  in professional sports.
\end{enumerate}

\noindent This works contributes to an emerging literature providing a complex systems perspective  on  sports competitions and people analytics, in particular by highlighting the remarkable level of variation in annual and career performance metrics. Such high levels of  variability point to the pervasive role of non-linear dynamics underlying the evolution of both individual and team competition.

  \vspace{-0.2in}
\section*{Methods}
\vspace{-0.2in}
We define prowess as an individual player's ability to succeed in achieving a specific outcome $x$ (e.g. a HR in MLB or a Reb. in the NBA) in any given opportunity $y$ (here defined to be an at-bat (AB) or Inning-Pitched-in-Outs (IPO), for batters and pitchers respectively, in MLB; or a minute played in the NBA). Thus, our method implicitly accounts for a fundamental source of variation over time, which is growth in league size and games per season, since all outcome measures analyzed are considered on a per-opportunity basis.

\begin{figure}
\centering{\includegraphics[width=0.99\textwidth]{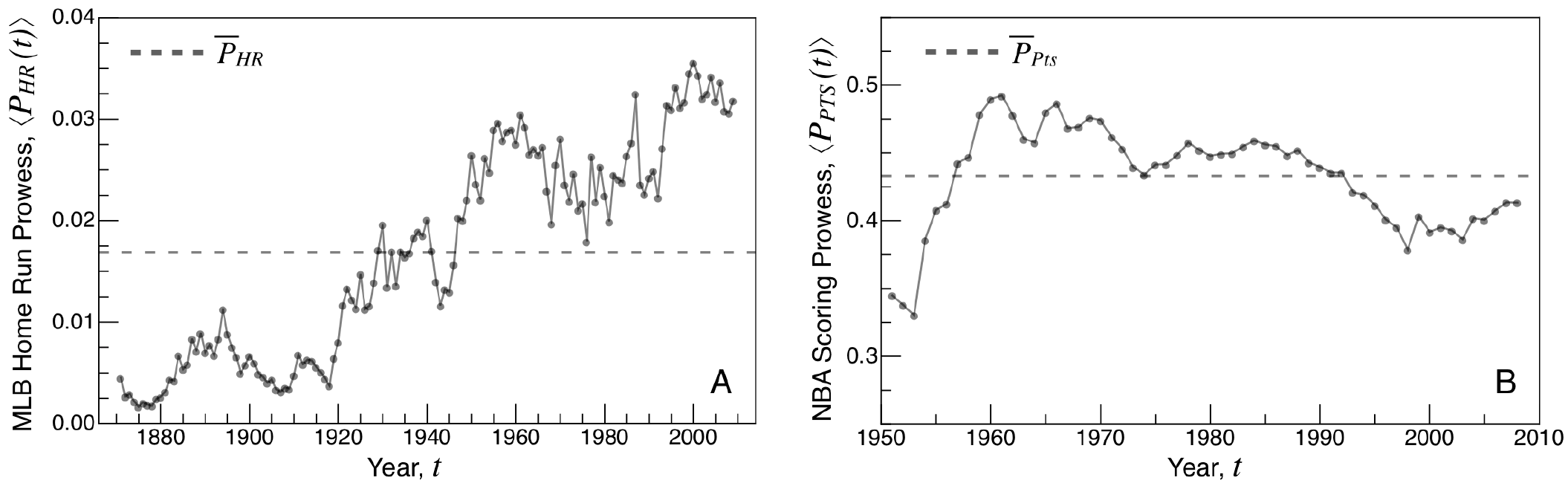}}
\caption{ \label{figure:F1} {\bf Non-stationary evolution of player prowess in professional Baseball and  Basketball.} The seasonal prowess $\langle P(t) \rangle$  measures the relative success  per opportunity rate using appropriate measures for a given sport. By normalizing accomplishments with respect to $\langle P(t) \rangle$, we objectively account for  variations in prowess derived from  endogenous and exogenous factors associated with the evolution of each sport. (A) The home-run prowess  shows a significant increasing trend since 1920, reflecting the emergence of the modern ``slugger'' in MLB. Physiological, technological, economic, demographic and social factors have played significant roles in MLB history \cite{ward1996baseball}, and are  responsible for sudden upward shifts observed for   $\langle P_{HR}(t) \rangle$. (B) Scoring prowess exhibits a non-monotonic trend.  Horizontal dashed lines correspond to the average value of each curve, $\overline{P}$, calculated over the entire period shown. See subpanels in Figure 4 for the prowess time series calculated for all 7 metrics analyzed. } 
\end{figure}

Figure \ref{figure:F1} shows the evolution of   home run prowess in MLB over the 139-year period 1871-2009 and the evolution of scoring prowess in the NBA over the 58-year period 1951-2008. It was beyond the scope of our analysis to update the performance data to present time, which is a clear limitation of our analysis, but such a right censoring issue is unavoidable with every passing year. Regardless, with data extending to the beginning of each league, our analysis accounts for several major paradigm shifts in each sport that highlight the utility of the method.
Indeed, while HR prowess has increased in era-specific bursts, point-scoring prowess shows different non-monotonic behavior that peaked in the early 1960s. Taken together, these results demonstrate the non-stationary evolution of player prowess over time with respect to the specific achievement metrics.
What this means from practical game, season and career perspectives,  is that the occurrence of a home run in 1920 was much  more significant from a statistical perspective (as it was relatively rarer per opportunity) than  a home run at the turn of the 21st century, which was  was the peak period of HR prowess (during  which numerous players were implicated by the Mitchell Report \cite{mitchell2007report} regarding an investigation into performance-enhancing drug sue in MLB). By way of economic analogy,  while the nominal baseball ticket price  in the early 20th century was around 50 cents,  the same ticket  price might  nominally be 100 times as much in present day USD\$, which points to the classic problem of comparing crude nominal values. To address this measurement problem, economists developed the  `price deflator' to  account for the discrepancy in nominal values by mapping  values recorded in different periods to their `real' values, a procedure that requires measuring price values relative to a common baseline year. Hence, in what follows, our approach is a generalization of the common method used in economics  to account for long-term inflation, and readily extends to other metric-oriented domains biased by persistent secular growth, such as scientometrics  \cite{petersen_citationinflation_2018,petersen2018mobility}. 

Thus, here the average prowess serves as a baseline `deflator index' for comparing accomplishments achieved in different years and thus  distinct historical eras. We conjecture that the changes in the average prowess are related to league-wide factors which can be quantitatively neutralized (also referred to as  `detrended'  or `deflated') by renormalizing individual accomplishments by the average prowess for a given season.  To achieve this renormalization we first calculate the prowess $P_{i}(t)$ of an individual player $i$ as $P_{i}(t) \equiv x_{i}(t) / y_{i}(t)$, where $x_{i}(t)$  is an individual's total number of successes out of his/her total number $y_{i}(t)$  of opportunities in a given year $t$. 

To compute the league-wide average prowess, we then  compute the aggregate prowess as the success rate  across all opportunities, 
 \begin{equation}
 \langle P(t) \rangle \equiv \frac{\sum_{i} x_{i}(t)}{\sum_{i} y_{i}(t)}  \ .
 \end{equation}
In practical terms, we apply the summation across $i$ only over players with at least $y_c$ opportunities during year $t$; as such, the denominator represents the total number of opportunities across the subset of $N_{c}(t)$ players in year $t$. We implemented thresholds of $y_{c} \equiv$ 100 AB (batters), 100 IPO (pitchers), and 24 Min. (basketball players) to discount statistical fluctuations arising from players with very short seasons. The results of our renormalization method are robust to reasonable choices of  $y_{c}$ that exclude primarily just the  trivially short seasons, with a relatively large subset of $N_{c}(t)$ players remaining.

Finally, the renormalized achievement metric for player $i$ in year $t$ is given by
\begin{equation}
 x^{D}_{i}(t) \equiv x_{i}(t) \ \frac  {P_{\text{baseline}}}{\langle P(t) \rangle} \ ,
 \label{xdsingle}
\end{equation}
where $P_{\text{baseline}}$ is the arbitrary  value applied to all $i$ and all $t$, which  establishes a common baseline.
For example, in prior work \cite{petersen2011methods} we used  $P_{\text{baseline}} \equiv\overline{P}$, the average prowess calculated across  all years (corresponding to the dashed horizontal lines in Fig. \ref{figure:F1}).  Again, because the choice of baseline  is arbitrary, in this work we renormalize   HR statistics in MLB relative to the most recent prowess value, $P_{\text{baseline}} \equiv \langle P(2009) \rangle$,  a choice that facilitates contrasting with the results reported in \cite{petersen2011methods}; and for all 6 other performance metrics we normalize using $P_{\text{baseline}} \equiv \overline{P}$.

Applying this method we calculated renormalized metrics at both the single season level, corresponding to $x_{i}^{D}(t)$, and the total career level, corresponding to the aggregate player tally  given by
\begin{eqnarray}
 X^{D}_{i} &=& \sum_{s=1}^{L_{i}}x^{D}_{i}(s)  \ ,
 \label{XD}
  \end{eqnarray}
where  $s$ is an index for player season and $L_{i}$ is the player's career length measured in seasons.

\vspace{-0.2in}
\section*{Results}
\vspace{-0.2in}
We applied our renormalization method to two prominent and historically relevant North American professional sports leagues, using comprehensive  player data comprised of roughly 17,000 individual  careers spanning more than a century of league play in the case of MLB (1871-2009) and roughly 4,000 individual  careers spanning more than a half-century (1946-2008) of league play in the case of NBA. Together, these data represent roughly 104,000 career years and millions of in-game opportunities consisting of more than 13.4 million at-bats and 10.5 million innings-pitched-in-outs in the MLB, and 24.3 million minutes played in the NBA through the end of the 2000s decade.

\begin{figure}
\centering{\includegraphics[width=0.75\textwidth]{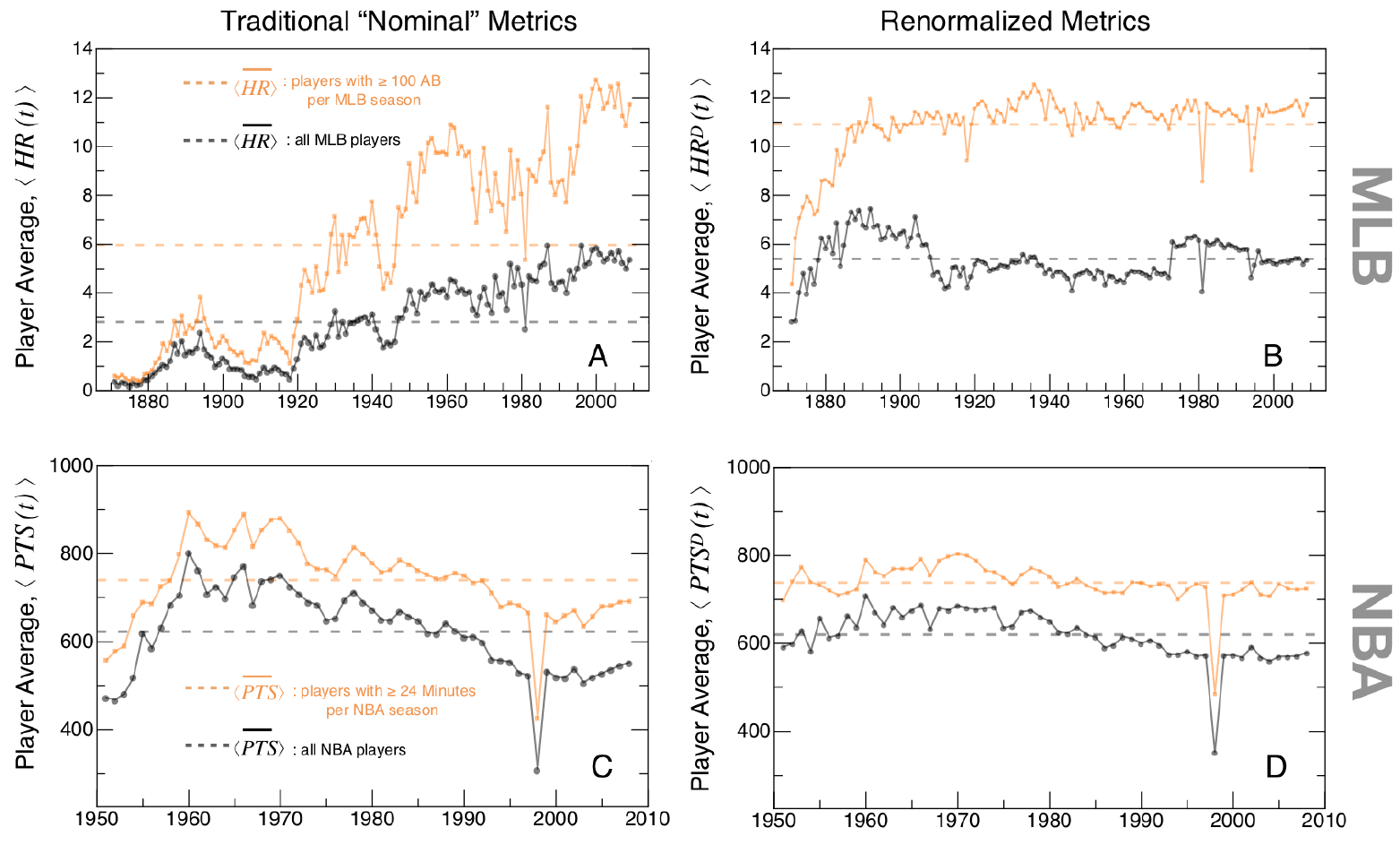}}
\caption{ \label{figure:F2} {\bf Renormalizing player performance metrics addresses systematic  performance-inflation bias.} The annual league average for  HR (MLB) and Pts. (NBA)   calculated before versus after  renormalizing the performance metrics -- i.e., panels (A,C) show $\langle x(t )\rangle$ and (B,D) show $\langle x^{D}(t )\rangle$.  League averages are calculated using all players (black) and a subset of players with sufficient season lengths as to avoid fluctuations due to those players with trivially short season lengths (orange). Horizontal dashed lines correspond to the average value of each curve over the entire period. Significant dips are due to prominent player strikes resulting in game cancellations,  in 1981, 1994 and 1995 for MLB and 1995-96 for the NBA.} 
\end{figure}

\begin{figure}
\centering{\includegraphics[width=0.75\textwidth]{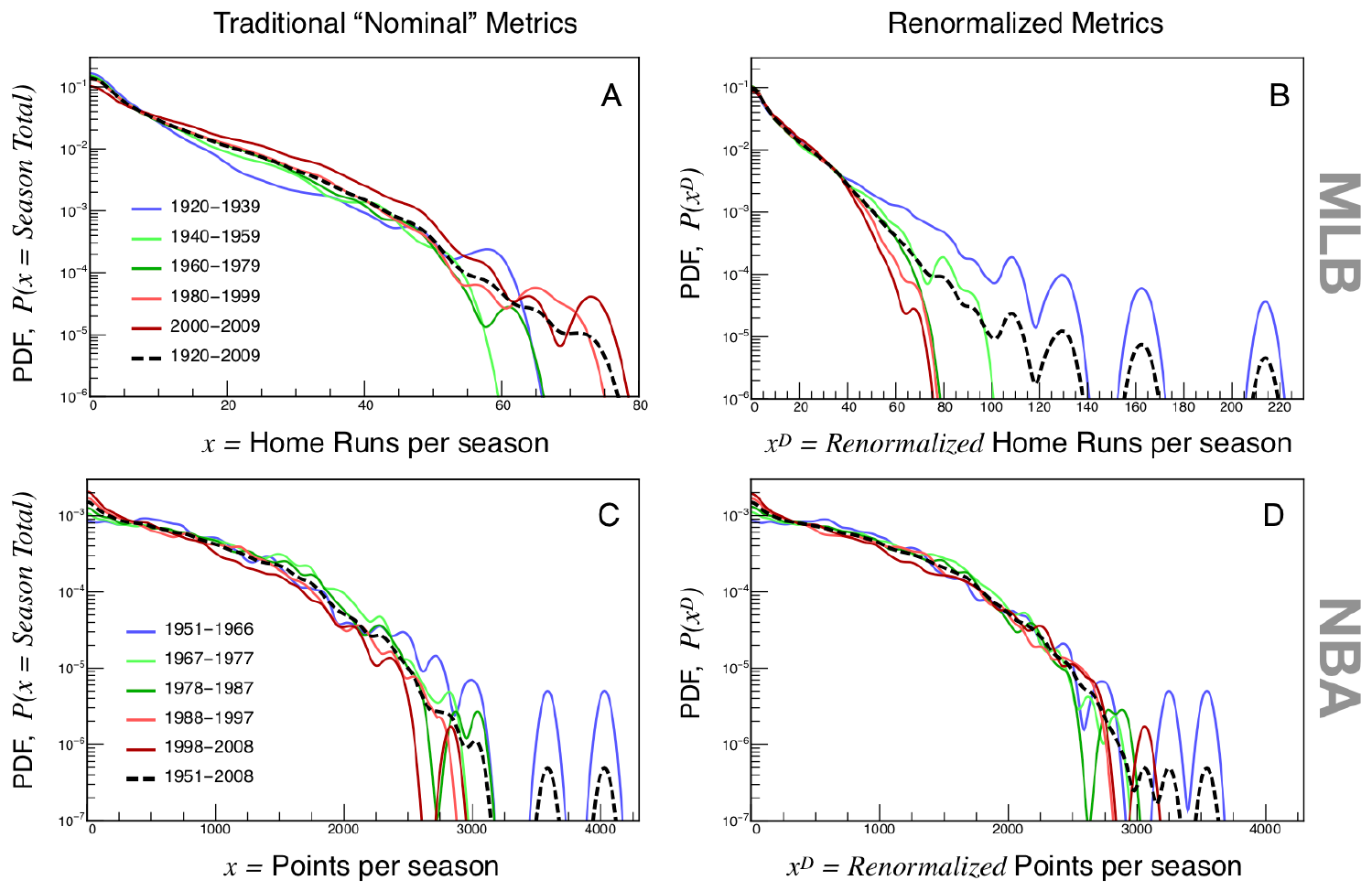}}
\caption{ \label{figure:F3} {\bf Distribution of annual player performance -- comparing traditional  and renormalized metrics.} Each curve corresponds to the distribution $P(x)$ for traditional (A,C) and renormalized metrics (B,D); season level data were separated into non-overlapping observation periods  indicated in each legend. $P(x)$  estimated using a kernel density approximation, which  facilitates identifying outlier values. The renormalized metrics in panels (B,D) show improved data collapse towards a common distribution for a larger range of $x$ values (but not including the extreme tails which correspond to outlier achievements), thereby confirming that our method facilitates the distillation of a universal distribution of seasonal player achievement. Regarding the case of HR in panels (A,B), our method facilitates highlighting outlier achievements that might otherwise be obscured by underlying shifts in prowess; such is the case for Babe Ruth's career years during the 1920's, which break the {\it all-time scales}, as shown in Appendix Table S1.} 
\end{figure}

Figure \ref{figure:F2} compares the league averages   for home runs in MLB and points scored in the NBA,  calculated using all  players in each year (black curves) and just the subset of players with $y_{i}(t) \geq y_{c}$ (orange curves) in order to demonstrate the robustness of the method with respect to the choice of $y_{c}$. More specifically,  Fig. \ref{figure:F2}(A,C) shows the league average based upon the traditional ``nominal'' metrics, computed as  $\langle x (t) \rangle \equiv N_{c}(t)^{-1}\sum_{i}x_{i}(t)$, while Fig. \ref{figure:F2}(B,D) show the league average based upon renormalized metrics, $\langle x(t)^{D} \rangle \equiv N_{c}(t)^{-1}\sum_{i}x^{D}_{i}(t)$; the sample size $N_{c}(t)$ counts the number of players per season satisfying the opportunity threshold  $y_{c}$. 

 In order to demonstrate the utility of this method to address the non-stationarity in the nominal or ``raw'' player data, we applied the Dickey-Fuller test \cite{dickey1979distribution} to the historical time series for   in per-opportunity success rates (measured by $\langle P(t) \rangle$) and the corresponding league averages ($\langle x(t )\rangle$ and $\langle x^{D}(t )\rangle$). More specifically, we applied the test using an autoregressive model with drift  to each player metric, and repoort the  test statistic and corresponding   $p$-value used to test the null hypothesis that the data follows a non-stationary process. For example, in the case of Home Runs: for the time series $\langle P(t) \rangle$ (respectively $\langle HR(t)\rangle$) we obtain a test statistic = -3.7 (-6.5) and corresponding $p$-value = 0.57 (0.3), meaning that we fail to reject the null hypothesis, thereby indicating that the prowess time series (league average time series) is non-stationary; contrariwise, for the renormalized league average time series $\langle HR^{D}(t)\rangle$ we obtain a test statistic = -31.5 and $p-$value = 0.0004 indicating that the data follow a stationary time series. Repeating the same procedure for Points:   for $\langle P(t) \rangle$ (resp. $\langle PTS(t)\rangle$) we obtain a test statistic = -6.4 (-9.6) and corresponding $p$-value = 0.3 (0.13), also  indicating that both time series are non-stationary; contrariwise, for the renormalized league average $\langle PTS^{D}(t)\rangle$ we obtain a test statistic = -22.2 and $p-$value = 0.003 indicating that the renormalized data follow a stationary time series. We observed this similar pattern, in which the renormalization method transforms non-stationary time series into a stationary time series, for Strikeouts, Rebounds and Assists; whereas in the case of Wins and Hits, the Dickey-Fuller test applied to $\langle P(t) \rangle$), $\langle x(t )\rangle$ indicate that these time series are already stationary.

Notably, as a result and consistent with a stationary data generation process, the  league averages are more constant over time after renormalization, thereby demonstrating the utility of this renormalization methods to standardize multi-era individual achievement metrics. Nevertheless, there remain deviations from a perfectly horizontal line following from phenomena not perfectly captured by our simple renormalization. Indeed, extremely short seasons and careers, and phenomena underlying the prevalence of these short careers, can bias the league average estimates. For example, in 1973 the designated hitter rule was introduced into the American League, comprising half of all MLB teams, which skews the number of at-bats per player by position, since half of pitchers no longer tended to take plate appearances after this rule change. Consequently, there is a prominent increase  in the average home runs per player in 1973 corresponding to this rule change, visible in  Figure \ref{figure:F2}(B) in the curve calculated for all MLB players (black curve), because roughly 1 in 2 pitchers (who are not typically power hitters) did not enter into the analysis thereafter. For similar reasons, our method does not apply as well to pitcher metrics because of a compounding decreasing trend in the average number of innings pitched per game due to the increased role of relief pitchers in MLB over time; accounting for such strategic shifts in the role and use of individual player types could also be included within our framework, but is outside the scope of the present discourse and so we leave it for future work.  See Fig. 5 in ref. \cite{petersen2011methods} for additional details regarding this detail, in addition to a more detailed development of our renormalization method in the context of MLB data only.

Based upon the convergence of the seasonal player averages to a consistent value that is weakly dependent on year, the next question is to what degree do the annual distributions for these player metrics collapse onto a common curve, both before and after application of the renormalization method.
To address this question,  Figure \ref{figure:F3} shows the probability density function (PDF) $P(x)$ for the same two metrics, HR and points scored, measured at the season level. 
For each case we separate the data into  several  non-overlapping periods. It is important to recall that  $P_{\text{baseline}} \equiv \langle P(2009) \rangle$ for HR and  $P_{\text{baseline}} \equiv \overline{P}$ for Pts. Consequently, there is a significant shift in the range of values for HR but not for Pts., which facilitates contrasting  the benefits provided by these two options. 

In the case of HR, the scale shifts from a maximum of 73 HR (corresponding to Barry Bond's 2001 single-season record) to 214 renormalized HR (corresponding to Babe Ruth's 59 nominal HR in 1921). While this latter value may be unrealistic, it nevertheless highlights the degree to which Babe Ruth's slugging achievements were outliers relative to his contemporaneous peers, further emphasizing the degree to which such achievements are  under-valued by comparisons based on  nominal metrics.  In the case of Pts., in which there is negligible rescaling due to the choice of $P_{\text{baseline}} \equiv \overline{P}$, we observe a compacting at the right tail rather than the divergence observed for HR.  And in both cases, we  observe a notable data collapse in the bulk of $P(x)$. For example, Fig. \ref{figure:F3}(B) collapses to a common  curve for the majority of the data, up to the level of $x^{D}\approx 35$ renormalized HR. In the case of NBA points scored, the data collapse  in Fig. \ref{figure:F3}(D)  extends to the level of $x^{D}\approx 2500$ renormalized Pts., whereas for the traditional metrics in  Fig. \ref{figure:F3}(C)  the data collapse extends to the $x^{D}\approx 1000$ renormalized Pts. level. 

\begin{figure}
\centering{\includegraphics[width=0.84\textwidth]{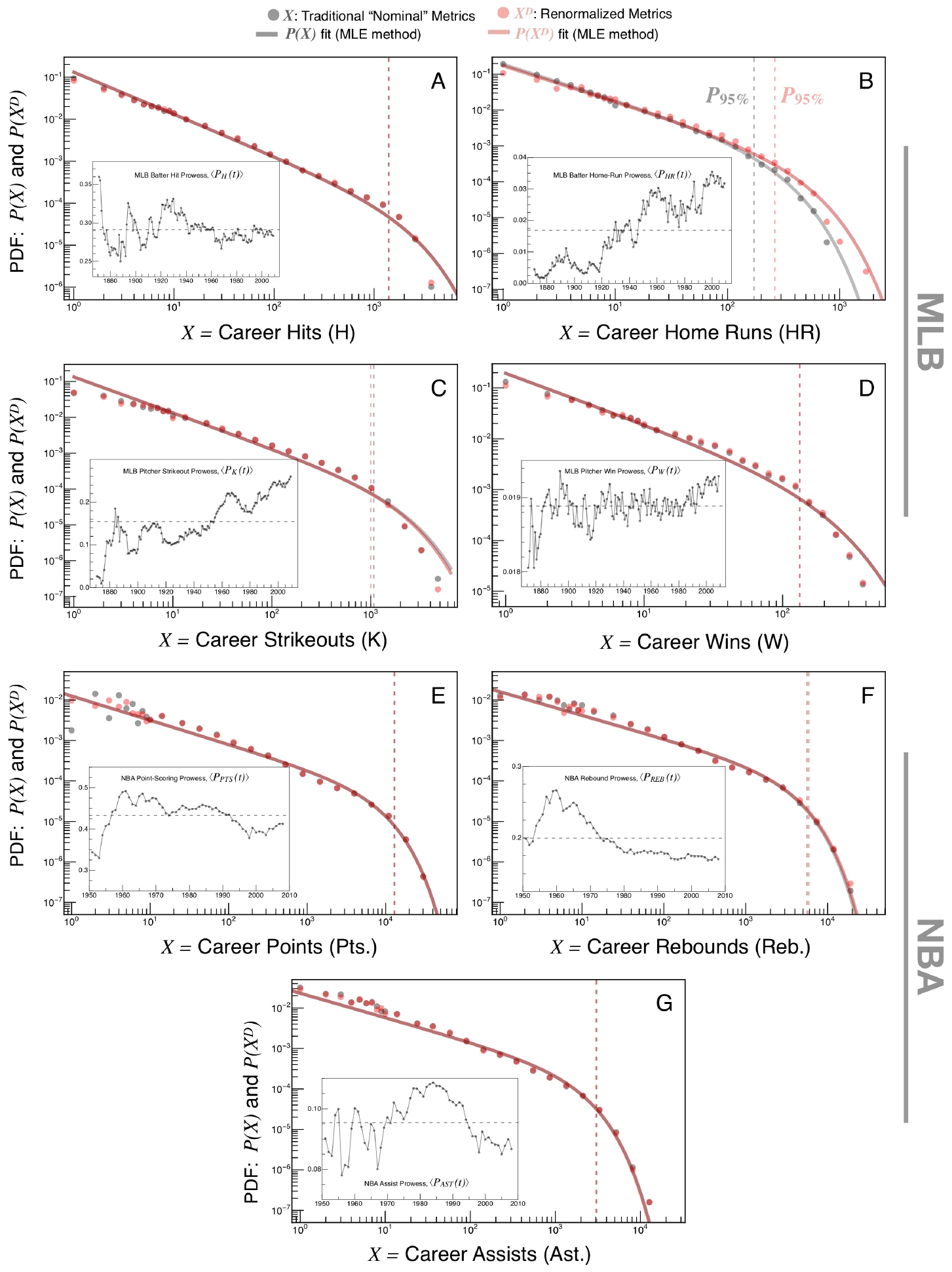}}
\caption{ \label{figure:F4} {\bf Distribution of career achievement totals -- comparing traditional and renormalized metrics.} Data points represent the empirical  PDFs calculated for traditional (gray) and renormalized (red) metrics. Vertical dashed lines indicate the location of the 95th percentile value ($P_{95}$) for each distribution, indicating the onset of {\it all-time greats} likely to be honored in each league's Hall of Fame. Each solid line corresponds to a distribution fit estimated using the MLE method; panels (A-D) are fit using the Log-Series distribution defined in Eq. (\ref{PDFLS}) and (E-G) are fit using the Gamma distribution defined in Eq. (\ref{PDFGamma}); see ref. \cite{petersen2011methods} for estimation of the best-fit Gamma distribution parameters for MLB data. The deviations between $P(X)$ and $P(X^{D})$ are less pronounced than the counterparts $P(x)$ and $P(x^{D})$ calculated at the seasonal level, indicating that the overall distribution of career achievement is less sensitive to shifts in player prowess -- however, this statement does not necessarily apply to the {\it ranking} of individuals, which can differ remarkably between the traditional and renormalized metrics. (Insets) Time series of average league prowess for each metric to facilitate cross-comparison and to highlight the remarkable statistical regularity in the career achievement distributions despite the variability in player prowess across time; Horizontal dashed lines correspond to the average value $\overline{P}$ calculated over the entire period shown. } 
\end{figure}

Figure \ref{figure:F4} shows the empirical distributions $P(X)$ and $P(X^{D})$ for  career totals, addressing to what degree does renormalization of season-level metrics impact the achievement distributions at the career level. Also plotted along with each empirical PDF is the distribution  model fit calculated using the Maximum Likelihood Estimation (MLE) method. In previous work \cite{petersen2011methods} we highlighted the continuous-variable  Gamma distribution as a theoretical model, given by
\begin{eqnarray}
 P_{\Gamma}(X \vert \alpha, X_{c}) &\propto& X^{-\alpha} \exp[-X/X_{c}] \ .
 \label{PDFGamma}
  \end{eqnarray}
This distribution is characterized by two parameters: the scaling parameter $\alpha$ (empirically observed sub-linear values range between 0.4 and 0.7) captures the power-law decay, while the location parameter $X_{c}$ represents the onset of extreme outlier achievement terminated  by an exponential cutoff arising from finite size effects (finite season and career lengths); see ref. \cite{petersen2011methods} for estimation of the best-fit Gamma distribution parameters for MLB data.  
 
We also  highlight an alternative theoretical model given by the discrete-variable  Log-Series distribution,
\begin{eqnarray}
 P_{LS}(X \vert p) &\propto& p^{X}/X \  \approx X^{-1} \exp[-X/X_{c}] \ .
 \label{PDFLS}
  \end{eqnarray}
In particular, this model distribution is characterized by a single parameter $0 < p < 1$; for example, in the case of HR we estimate $p=0.996975$. In such a case where  $p \approx 1$ (hence $1-p \ll 1$) then   the approximation in Eq. (\ref{PDFLS}) follows, giving rise to the exponential cutoff value $X_{c} = 1/(1-p)$. A historical note, the Log-Series PDF was originally proposed in ecological studies \cite{FisherLogSeriesDist}.

In this work we find $P_{LS}(X)$ to provide a better fit than  $P_{\Gamma}(X)$ for the empirical career distributions for MLB data, but not for NBA data.  As such, the fit curves for MLB in Fig. \ref{figure:F4}(A-D) correspond to $P_{LS}(X)$, whereas the fit curves for the NBA in Fig. \ref{figure:F4}(E-F) correspond to $ P_{\Gamma}(X)$.    This subtle difference in the functional form of the $P(X)$ distributions may be the starting point for understanding variations in competition and career development between these two professional sports. We refer the detail-oriented reader to ref. \cite{petersen2011quantitative} for further discussion on the analytic properties of $ P_{\Gamma}(X \vert \alpha, X_{c})$, as derived from a theoretical model of career longevity, which provides an intuitive mechanistic understanding of $\alpha$ and $X_{c}$.  While in previous work we have emphasized the estimation, significance and meaning of distribution parameters, here we are motivated to demonstrate the generalizability of the renormalization method, and so we leave the analysis of different $P(X)$ parameter estimations between leagues as a possible avenue for future research. 

Notably, Figure \ref{figure:F4} shows that each pair of empirical data, captured by $P(X)$ and $P(X^{D})$, exhibit relatively small deviations from each other in distribution. 
Interestingly, metrics representing  achievements with relatively lower per-opportunity success rates per opportunity (home runs, strikeouts and rebounds) are more sensitive to time-dependent success factors than  those with relatively higher success rates (hits, wins and points). This pattern can also be explained in the context of the Dickey-Fuller test results, which indicated that Wins and Hits metrics are sufficiently stationary to begin with.
In all, our results indicate that  the extremely right-skewed  (heavy-tailed) nature of player achievement distributions reflect intrinsic properties underlying achievement that are robust to inflationary and deflationary factors that influence success rates, once accounted for. The stability of the $P(X)$ and $P(X^{D})$ distributions at the aggregate level is offset by the local reordering at the rank-order level -- see  Supplementary Material Appendix for ranked tables. In short,  for the NBA we provide 6 extensive tables   that list the top-50 all-time achievements comparing traditional and  renormalized metrics -- at both the season and career level; and for MLB we  provide a top-20 ranking for career home runs, and refer the curious reader to ref. \cite{petersen2011methods} for analog tables listing   top-50  rankings.

\vspace{-0.2in}
\section*{Discussion}
\vspace{-0.2in}
The analysis of career achievement  features many characteristics of generic multi-scale complex systems. For example, we document non-stationarity arising from the growth of the system along with sudden  shifts in player prowess following rule changes (i.e. policy interventions).  Other characteristics frequently encountered in complex systems are the entry and exit dynamics associated with finite life-course and variable career lengths, and memory with consequential path dependency   associated with cumulative advantage mechanisms underlying individual pathways to success. To address these challenges, researchers have applied concepts and methods from statistical physics \cite{petersen2008distribution,petersen2011quantitative,petersen2012persistence,schaigorodsky2014memory} and network science \cite{saavedra2010mutually,radicchi2011best,mukherjee2012identifying,mukherjee2014quantifying} to professional athlete data, revealing statistical regularities that provide a better understanding of the underlying nature of competition. Notably, academia also exhibits analogous statistical patterns that likely emerge from the general principles of competitive systems, such as  the extremely high barriers to entry which may explain the highly skewed career longevity distributions \cite{petersen2011quantitative} and first-mover advantage  dynamics that amplify the long-term impact of uncertainty \cite{petersen2012persistence}. By analogy,  renormalized  scientometrics   are  needed in order to compare researcher achievements across broad time periods \cite{petersen_citationinflation_2018}, for example recent work leveraged renormalized citations to   compare the effects of  researcher mobility  across a panel of individuals spanning several decades \cite{petersen2018mobility}.

Motivated by the application of complex systems science to the emerging domain of people analytics, we analyzed comprehensive player data from two prominent  sports leagues in order to objectively address a timeless question -- who's the greatest of all time? To this end, we applied our renormalization  method in order  to obtain performance metrics that are more suitable for cross-era comparison, thereby addressing motivation (1) identified in the introduction section. From a practical perspective, our method renormalizes player achievement metrics with respect to player success rates, which facilitates removing time-dependent trends in performance ability relating to various  physiological, technological, and economic factors. 
In particular, our method accounts for various types of historical  events  that  have increased  or decreased  the rates of success per player opportunity, e.g. modern  training regimens, PEDs, changes in the physical construction of bats and balls and shoes, sizes of ballparks, talent dilution of players from expansion, etc.  While in previous work we applied our renormalization method exclusively to MLB career data \cite{petersen2011methods}, here we demonstrate the generalizability of the method by applying it to  an entirely different sport. Since renormalized metrics facilitate  objective comparison of player achievements across distinct league eras, in principal an appropriate cross-normalization could also facilitate comparison across different sports.

The principal requirements of our renormalization method are: (a)  individual-oriented metrics recording achievements as well as opportunities, even if the sport is  team-oriented; and (b) data be    comprehensively available for all  player opportunities so that per-opportunity success rates can be consistently and robustly estimated. 
We then use the prowess time-series $\langle P(t) \rangle$ as an `achievement deflator' to robustly capture  time-dependent  performance factors. Take for example assists in the NBA, for which the average player prowess $\langle P(t) \rangle$ peaked in 1984 during the era of point-guard dominance in the NBA, and then decreased  25\% by 2008 (see Fig. \ref{figure:F4}G). This decline captures a confluence of factors including shifts in  team strategy and dynamics, as well as other individual-level factors  (i.e. since an assist is contingent on another player scoring, assist frequencies depend also on scoring prowess). More generally, such performance factors may affect   players differently depending on their team position or specialization, and so this is another reason why comprehensive player data is necessary to capture league-wide paradigm shifts. 

 The choice of renormalization baseline $P_{\text{baseline}}$ also  affects the resulting renormalized metric range. Consequently, the arbitrary value selected for $P_{\text{baseline}}$  can be used to emphasize the occasional apparently super-human achievements of foregone greats when measured  using contemporary metrics. For example, we highlight the ramification of this choice in the case of home runs, for which we used $P_{\text{baseline}} \equiv \langle P_{HR}(2009) \rangle$, such  that Fig. \ref{figure:F3}(B) shows season home-run tallies measured in units of 2009 home-runs. As a result, the maximum value in the season home-run distribution corresponds to Babe Ruth's career year in 1921 (and in fact not 1927, when HR prowess was relatively higher) in which he hit the equivalent of 214 renormalized Home Runs (or 2009 HRs). 
 Alternatively,  we also demonstrate how using the average prowess value as the  baseline,  $P_{\text{baseline}} \equiv \overline{P}$, yields a renormalized  metric range that is more consistent with the range of traditional metrics, as illustrated by the distributions in Fig. \ref{figure:F3}(C,D). In such  cases when the prowess time series is non-monotonic, there may not be a unique year corresponding to a given prowess value used as  $P_{\text{baseline}}$. This is the case for assists, see  Fig. \ref{figure:F4}(G),  since assist prowess peaked in the mid-1980s. As a result,  renormalized assist metrics for players significantly before or after this period, when prowess values were lower, will have relatively greater renormalized assist metrics.  
 
  To facilitate visual inspection of how the nominal values translate into renormalized values, we provide 6 tables in the Supplementary Material Appendix that rank NBA metrics at the  season and career levels (see \cite{petersen2011methods} for analog tables ranking MLB player achievements). 
 All tables are split into left (traditional ranking) and right sides (renormalized ranking). For example,  Table S6 starts with: \\
 
 \begin{table}[h!]
\begin{tabular}{lccc||lcccc}
&\multicolumn4c{{Traditional Rank}}&\multicolumn4c{{Renormalized Rank}}\\
 Rank & Name & Season ($Y\#$) & Season Metric & Rank$^{*} $(Rank)  &  \% Change & Name & Season ($Y\#$) & Season Metric \\
 \hline
1 & Wilt  Chamberlain & 1960  (2)  &2149 & 1(28)  & 96 & Dennis  Rodman & 1991  (6)  &1691 \\
\end{tabular}
\end{table}
\noindent This line indicates that  in  the 1960-61 season, Wilt Chamberlain obtained 2149 rebounds, the most for a single season, corresponding to his second career year (Y\#).
However, according to renormalized metrics,  Dennis Rodman's 6th career year in the 1991-1992 season finds new light as the greatest achievement in terms of renormalized rebounds (1691), despite being ranked \#28th all-time according to the nominal value (1530 rebounds), a shift corresponding  to a 96\% percent  rank increase. 
Not all metrics display such profound re-ranking among the all-time achievements. Such is the case for Wilt Chamberlain's single-season scoring record (see Table S5) and John Stockton's single-season assists record  (see Table S7), which maintain their top ranking after renormalization.
 
Also at the season level, another source of variation in addition to performance factors is the wide range of ability and achievement rates across individuals. 
Consequently, renormalization based upon average league prowess, $\langle P(t) \rangle$, can be strongly influenced by   outlier achievements at the player-season level.  Fig. \ref{figure:F3} illustrates season-level performance distributions for HR and Points, comparing the distributions calculated for nominal metrics, $P(x)$, and renormalized metrics, $P(x^{D})$. 
Because $\langle P(t) \rangle$ captures average performance levels, the data collapse across achievement distributions drawn from multiple eras in Fig. \ref{figure:F3} is weakest in the right tails that capture outlier player performance. Nevertheless, the data collapse observed in the bulk of the $P(x^{D})$ distributions indicates that the variation in player achievements, an appropriate proxy for league competitiveness, has been relatively stable over the history of each league. 

At the career level, this comprehensive study of all player careers facilitates a better appreciation for the relatively high frequencies of {\it one-hit wonders} -- individuals with nearly minimal achievement metrics -- along with much smaller but statistically regular and theoretically predictable frequencies  of superstar careers. By way of example, previous work reveals that roughly 3\% of non-pitchers (pitchers) have a career lasting only one at-bat (lasting an inning or less) and 5\% of non-pitchers complete their  career with just a single hit; Yet,  the same profession also sustains careers that span more than 2,000 games, 10,000 at bats and 4,000 innings pitched \cite{petersen2011methods}. Here we find that the same disparities hold for players in a different sport with different team dynamics, player-player interactions, and career development system  (e.g. the NBA  introduced a `minor league' system in 2001). In particular, 3\% of NBA careers end within the first 1-12 minutes played, and 2\% of careers last only 1 game! Yet, the average career length is roughly 273 games (roughly 3 seasons), while the maximum career length is owed to Robert Parish with 1,611 games, almost six times the average. Another anomaly is Kareem Abdul-Jabbar's career, which spanned 57,446 minutes played, roughly 9 times the average career length measured in minutes. Similar results have also been observed for professional tennis careers \cite{radicchi2011best}. Such comparisons between extreme achievers and average player performance illustrate the difficulty in defining  a `typical' player  in light of such right-skewed achievement distributions.
This lack of characteristic scale is evident in the career achievement distributions shown in Fig. \ref{figure:F4}, which indicate a  continuum of achievement totals across the entire range. In other words, these professional sport leagues breed one-hit wonders, superstars and all types of careers inbetween -- following a highly regular statistical pattern that bridges the gap between the extremes.

Remarkably, Fig. \ref{figure:F4} indicates little variation when comparing the career achievement distribution $P(X)$ calculated using traditional metrics against the corresponding $P(X^{D})$ calculated using renormalized career metrics. 
This observation provides several insights and relevant policy implications. First, the invariance indicates that the extremely right skewed distribution of career achievement are not merely the result of mixed era-specific distributions characterized by different parameters and possibly different functional forms. Instead, this stability points to a  universal distribution of career achievement that likely follows from simple parsimonious system dynamics. Second, this invariance also indicates  that the all-time greats were not {\it born on another planet}, but rather, follow naturally form the statistical regularity observed in the player achievement distributions, which feature common lower  and upper tail behavior representing the most common and most outstanding careers, respectively.  
Third, considering benchmark achievements in various sports, such as the 500 HR  and 3000 K clubs in MLB, and the 20,000 points, 10,000 rebounds and 5,000 assists clubs in the NBA, such invariance indicates that such thresholds are nevertheless  stable with respect to the time-dependent factors where renormalized metrics are used.
This latter point follows because, while the distribution may be stable, the ranking of individuals is not. Such local rank-instability provides additional fodder for casual argument and serious consideration among fans and statisticians alike.
And finally, regarding the preservation of cultural heritage, these considerations can be informative to both Baseball and Basketball Hall of Fame selection committees, in particular  to address motivation (2) identified in the introduction section concerning  standard and retroactive player induction.\\

\noindent{\bf Acknowledgements:} We are indebted to countless sports \& data fanatics who helped to compile the comprehensive player data, in particular Sean Lahman \cite{BaseballData}. We are also grateful to two reviewers for their insightful comments.


\clearpage
\newpage

\beginsupplement

\begin{center}
{\bf \Large Supplementary Material Appendix: \\
\bigskip
\bigskip
Renormalizing individual performance metrics for cultural heritage management of sports records\\
 \bigskip
 \bigskip
Tables S1-S7} \\
 \bigskip
 \bigskip
 \bigskip
 \bigskip
 \bigskip
\bigskip
Alexander M. Petersen$^{1}$, Orion Penner$^{2}$, \\ 
\bigskip
$^{1}$ 
 Management of Complex Systems Department,\\
 Ernest and Julio Gallo Management Program,\\
School of Engineering, University of California, Merced, CA 95343  \\
$^{2}$ Chair of Innovation and Intellectual Property Policy, College of Management of Technology,\\
 Ecole Polytechnique Federale de Lausanne, Lausanne, Switzerland.  \\
\end{center}

Renormalized metrics are calculated using $P_{\text{baseline}} \equiv \overline{P}$ in   Tables S1-S7.
In Table S1 we list top-20 rankings for MLB home runs; see the Online Supplementary Information for Petersen, Penner and  Stanley, EPJB 2011 \cite{petersen2011methods} for additional top-50 rankings for MLB.
In Tables S2-S7 we list top-50 rankings for NBA metrics, including points, assists and rebounds over the career and for individual seasons. For the two types of rankings, career and season, the columns are organized as follows:\\

{\bf Career Tables S1--S4}: The 4 columns on the left of each table list information for the ``traditional rank'' of career
statistics, where the top 50 players are ranked along with their final season (career length in seasons listed in
parenthesis) and their career metric tally. The 5  columns on the right of each table list information for the ``renormalized rank'' ($Rank^{*}$) of career statistics, where the corresponding traditional rank (Rank) of the player is denoted in parenthesis.
 $L$ denotes the career length of the player.  The relative percent change  $\% Change = 100 (Rank-Rank^{*})/Rank$. \\
 
{\bf Season Tables S5--S7}: The 4 columns on the left  list the traditional ranking
of season statistics, where the top 50 players are ranked along with the year. The right columns list the renormalized
ranking of season statistics $Rank^{*}$. $Y\#$ denotes the number of years into the career. The relative percent change  $\% Change = 100(Rank-Rank^{*})/Rank$. 

\begin{table}
\centering{ {\footnotesize
\begin{tabular}{@{\vrule height .5 pt depth4pt  width0pt}lccc||lccc}
&\multicolumn4c{{\bf Traditional Rank}}&\multicolumn3c{{\bf Renormalized Rank}}\\
\noalign{
\vskip-1pt} Rank & Name & Final Season (L) & Career Metric & Rank$^{*}$(Rank)  &  Name & Final Season (L)
& Career Metric \\
\hline 
1 & Barry  Bonds & 2007  (22)  &762 & 1(3)  &  Babe  Ruth & 1935  (22)  &1215 \\
2 & Hank  Aaron & 1976  (23)  &755 & 2(23)  &  Mel  Ott & 1947  (22)  &637 \\
3 & Babe  Ruth & 1935  (22)  &714 & 3(26)  &  Lou  Gehrig & 1939  (17)  &635 \\
4 & Willie  Mays & 1973  (22)  &660 & 3(17)  &  Jimmie  Foxx & 1945  (20)  &635 \\
5 & Ken  Griffey Jr. & 2009  (21)  &630 & 5(2)  &  Hank  Aaron & 1976  (23)  &582 \\
6 & Sammy  Sosa & 2007  (18)  &609 & 6(124)  &  Rogers  Hornsby & 1937  (23)  &528 \\
7 & Frank  Robinson & 1976  (21)  &586 & 7(192)  &  Cy  Williams & 1930  (19)  &527 \\
8 & Alex  Rodriguez & 2009  (16)  &583 & 8(1)  &  Barry  Bonds & 2007  (22)  &502 \\
8 & Mark  McGwire & 2001  (16)  &583 & 9(4)  &  Willie  Mays & 1973  (22)  &490 \\
10 & Harmon  Killebrew & 1975  (22)  &573 & 10(18)  &  Ted  Williams & 1960  (19)  &482 \\
11 & Rafael  Palmeiro & 2005  (20)  &569 & 11(13)  &  Reggie  Jackson & 1987  (21)  &478 \\
12 & Jim  Thome & 2009  (19)  &564 & 12(14)  &  Mike  Schmidt & 1989  (18)  &463 \\
13 & Reggie  Jackson & 1987  (21)  &563 & 13(7)  &  Frank  Robinson & 1976  (21)  &444 \\
14 & Mike  Schmidt & 1989  (18)  &548 & 14(10)  &  Harmon  Killebrew & 1975  (22)  &437 \\
15 & Manny  Ramirez & 2009  (17)  &546 & 15(577)  &  Gavvy  Cravath & 1920  (11)  &433 \\
16 & Mickey  Mantle & 1968  (18)  &536 & 16(718)  &  Honus  Wagner & 1917  (21)  &420 \\
17 & Jimmie  Foxx & 1945  (20)  &534 & 17(18)  &  Willie  McCovey & 1980  (22)  &417 \\
18 & Ted  Williams & 1960  (19)  &521 & 18(557)  &  Harry  Stovey & 1893  (14)  &413 \\
18 & Frank  Thomas & 2008  (19)  &521 & 19(5)  &  Ken  Griffey Jr. & 2009  (21)  &411 \\
18 & Willie  McCovey & 1980  (22)  &521 & 20(28)  &  Stan  Musial & 1963  (22)  &410 \\
\hline
\end{tabular}}}
\caption{  Ranking of Career Home Runs (1871 - 2009). 
{\scriptsize The left columns lists the traditional ranking of career
statistics, where the top 20 players are ranked along with their final season (career length in seasons listed in
parenthesis) and their career metric tally. The right columns list the renormalized ranking of career statistics
$Rank^{*}$, where the corresponding traditional ranking of the player is denoted in parenthesis.
 $L$ denotes the career length of the player.   In contrast to the main manuscript, in order to facilitate more intuitive  comparison, renormalized HR metrics reported in this table are calculated using $P_{\text{baseline}} \equiv \overline{P}$ rather than $P_{\text{baseline}} \equiv P(2009)$.}
 }
\label{table:careerHR}
\end{table}

\begin{table}[h]
\centering{ {\footnotesize
\begin{tabular}{@{\vrule height .5 pt depth4pt  width0pt}lccc||lcccc}
&\multicolumn4c{{\bf Traditional Rank}}&\multicolumn4c{{\bf Renormalized Rank}}\\
\noalign{
\vskip-1pt} Rank & Name & Final Season (L) & Career Metric & Rank$^{*}$(Rank)  &  \% Change & Name & Final Season (L)
& Career Metric \\
\hline 
1 & Kareem  Abdul-jabbar & 1988  (20)  &38387 & 1(2)  & 50 & Karl  Malone & 2003  (19)  &38033 \\
2 & Karl  Malone & 2003  (19)  &36928 & 2(1)  & -100 & Kareem  Abdul-jabbar & 1988  (20)  &36687 \\
3 & Michael  Jordan & 2002  (15)  &32292 & 3(3)  & 0 & Michael  Jordan & 2002  (15)  &32511 \\
4 & Wilt  Chamberlain & 1972  (14)  &31419 & 4(7)  & 42 & Shaquille  O'neal & 2008  (17)  &29575 \\
5 & Julius  Erving & 1986  (16)  &30026 & 5(5)  & 0 & Julius  Erving & 1986  (16)  &28934 \\
6 & Moses  Malone & 1994  (21)  &29580 & 6(4)  & -50 & Wilt  Chamberlain & 1972  (14)  &28615 \\
7 & Shaquille  O'neal & 2008  (17)  &27619 & 7(6)  & -16 & Moses  Malone & 1994  (21)  &28532 \\
8 & Dan  Issel & 1984  (15)  &27482 & 8(10)  & 20 & Hakeem  Olajuwon & 2001  (18)  &27177 \\
9 & Elvin  Hayes & 1983  (16)  &27313 & 9(8)  & -12 & Dan  Issel & 1984  (15)  &26362 \\
10 & Hakeem  Olajuwon & 2001  (18)  &26946 & 10(16)  & 37 & Reggie  Miller & 2004  (18)  &26361 \\
11 & Oscar  Robertson & 1973  (14)  &26710 & 11(12)  & 8 & Dominique  Wilkins & 1998  (15)  &26110 \\
12 & Dominique  Wilkins & 1998  (15)  &26668 & 12(21)  & 42 & Allen  Iverson & 2008  (13)  &26040 \\
13 & George  Gervin & 1985  (14)  &26595 & 13(9)  & -44 & Elvin  Hayes & 1983  (16)  &26035 \\
14 & John  Havlicek & 1977  (16)  &26395 & 14(22)  & 36 & Kobe  Bryant & 2008  (13)  &25797 \\
15 & Alex  English & 1990  (15)  &25613 & 15(13)  & -15 & George  Gervin & 1985  (14)  &25666 \\
16 & Reggie  Miller & 2004  (18)  &25279 & 16(20)  & 20 & Patrick  Ewing & 2001  (17)  &25129 \\
16 & Rick  Barry & 1979  (14)  &25279 & 17(14)  & -21 & John  Havlicek & 1977  (16)  &24796 \\
18 & Jerry  West & 1973  (14)  &25192 & 18(15)  & -20 & Alex  English & 1990  (15)  &24551 \\
19 & Artis  Gilmore & 1987  (17)  &24941 & 19(11)  & -72 & Oscar  Robertson & 1973  (14)  &24459 \\
20 & Patrick  Ewing & 2001  (17)  &24815 & 20(19)  & -5 & Artis  Gilmore & 1987  (17)  &24023 \\
21 & Allen  Iverson & 2008  (13)  &23983 & 21(16)  & -31 & Rick  Barry & 1979  (14)  &23893 \\
22 & Kobe  Bryant & 2008  (13)  &23820 & 22(23)  & 4 & Charles  Barkley & 1999  (16)  &23748 \\
23 & Charles  Barkley & 1999  (16)  &23757 & 23(28)  & 17 & Gary  Payton & 2006  (17)  &23374 \\
24 & Robert  Parish & 1996  (21)  &23334 & 24(18)  & -33 & Jerry  West & 1973  (14)  &23115 \\
25 & Adrian  Dantley & 1990  (15)  &23177 & 25(31)  & 19 & Kevin  Garnett & 2008  (14)  &23111 \\
26 & Elgin  Baylor & 1971  (14)  &23149 & 26(24)  & -8 & Robert  Parish & 1996  (21)  &22615 \\
27 & Clyde  Drexler & 1997  (15)  &22195 & 27(25)  & -8 & Adrian  Dantley & 1990  (15)  &22230 \\
28 & Gary  Payton & 2006  (17)  &21813 & 28(27)  & -3 & Clyde  Drexler & 1997  (15)  &22035 \\
29 & Larry  Bird & 1991  (13)  &21791 & 29(34)  & 14 & David  Robinson & 2002  (14)  &21578 \\
30 & Hal  Greer & 1972  (15)  &21586 & 30(38)  & 21 & Ray  Allen & 2008  (13)  &21338 \\
31 & Kevin  Garnett & 2008  (14)  &21277 & 31(35)  & 11 & Mitch  Richmond & 2001  (14)  &21192 \\
32 & Walt  Bellamy & 1974  (14)  &20941 & 32(26)  & -23 & Elgin  Baylor & 1971  (14)  &21163 \\
33 & Bob  Pettit & 1964  (11)  &20880 & 33(29)  & -13 & Larry  Bird & 1991  (13)  &20946 \\
34 & David  Robinson & 2002  (14)  &20790 & 34(44)  & 22 & Tim  Duncan & 2008  (12)  &20921 \\
35 & Mitch  Richmond & 2001  (14)  &20497 & 35(40)  & 12 & Clifford  Robinson & 2006  (18)  &20726 \\
36 & Tom  Chambers & 1997  (16)  &20049 & 36(47)  & 23 & Dirk  Nowitzki & 2008  (11)  &20641 \\
37 & John  Stockton & 2002  (19)  &19711 & 37(54)  & 31 & Paul  Pierce & 2008  (11)  &20204 \\
38 & Ray  Allen & 2008  (13)  &19661 & 38(37)  & -2 & John  Stockton & 2002  (19)  &20203 \\
39 & Bernard  King & 1992  (14)  &19655 & 39(33)  & -18 & Bob  Pettit & 1964  (11)  &20033 \\
40 & Clifford  Robinson & 2006  (18)  &19591 & 40(58)  & 31 & Vince  Carter & 2008  (11)  &19768 \\
41 & Walter  Davis & 1991  (15)  &19521 & 41(30)  & -36 & Hal  Greer & 1972  (15)  &19734 \\
42 & Terry  Cummings & 1999  (18)  &19460 & 42(50)  & 16 & Scottie  Pippen & 2003  (17)  &19595 \\
43 & Bob  Lanier & 1983  (14)  &19248 & 43(36)  & -19 & Tom  Chambers & 1997  (16)  &19433 \\
44 & Tim  Duncan & 2008  (12)  &19246 & 44(32)  & -37 & Walt  Bellamy & 1974  (14)  &19277 \\
45 & Eddie  Johnson & 1998  (17)  &19202 & 45(56)  & 19 & Glen  Rice & 2003  (15)  &19146 \\
46 & Gail  Goodrich & 1978  (14)  &19181 & 46(42)  & -9 & Terry  Cummings & 1999  (18)  &19051 \\
47 & Dirk  Nowitzki & 2008  (11)  &19084 & 47(49)  & 4 & Dale  Ellis & 1999  (17)  &18999 \\
48 & Reggie  Theus & 1990  (13)  &19015 & 48(39)  & -23 & Bernard  King & 1992  (14)  &18914 \\
49 & Dale  Ellis & 1999  (17)  &19002 & 49(71)  & 30 & Tracy  Mcgrady & 2008  (12)  &18808 \\
50 & Scottie  Pippen & 2003  (17)  &18940 & 50(41)  & -21 & Walter  Davis & 1991  (15)  &18748 \\
\hline
\end{tabular}}}
\caption{  Ranking of Career Points. 
{\scriptsize The left columns lists the traditional ranking of career
statistics, where the top 50 players are ranked along with their final season (career length in seasons listed in
parenthesis) and their career metric tally. The right columns list the renormalized ranking of career statistics
$Rank^{*}$, where the corresponding traditional ranking of the player is denoted in parenthesis. $L$ denotes the career
length of the player. } }
\label{table:Cpts}
\end{table}

\begin{table}[h]
\centering{ {\footnotesize
\begin{tabular}{@{\vrule height .5 pt depth4pt  width0pt}lccc||lcccc}
&\multicolumn4c{{\bf Traditional Rank}}&\multicolumn4c{{\bf Renormalized Rank}}\\
\noalign{
\vskip-1pt} Rank & Name & Final Season (L) & Career Metric & Rank$^{*}$(Rank)  &  \% Change & Name & Final Season (L)
& Career Metric \\
\hline 
1 & Wilt  Chamberlain & 1972  (14)  &23924 & 1(1)  & 0 & Wilt  Chamberlain & 1972  (14)  &19896 \\
2 & Bill  Russell & 1968  (13)  &21620 & 2(3)  & 33 & Moses  Malone & 1994  (21)  &19323 \\
3 & Moses  Malone & 1994  (21)  &17834 & 3(4)  & 25 & Kareem  Abdul-jabbar & 1988  (20)  &17782 \\
4 & Kareem  Abdul-jabbar & 1988  (20)  &17440 & 4(2)  & -100 & Bill  Russell & 1968  (13)  &17424 \\
5 & Artis  Gilmore & 1987  (17)  &16330 & 5(5)  & 0 & Artis  Gilmore & 1987  (17)  &16924 \\
6 & Elvin  Hayes & 1983  (16)  &16279 & 6(7)  & 14 & Karl  Malone & 2003  (19)  &16907 \\
7 & Karl  Malone & 2003  (19)  &14967 & 7(8)  & 12 & Robert  Parish & 1996  (21)  &16178 \\
8 & Robert  Parish & 1996  (21)  &14715 & 8(6)  & -33 & Elvin  Hayes & 1983  (16)  &16136 \\
9 & Nate  Thurmond & 1976  (14)  &14464 & 9(12)  & 25 & Hakeem  Olajuwon & 2001  (18)  &15463 \\
10 & Walt  Bellamy & 1974  (14)  &14241 & 10(13)  & 23 & Buck  Williams & 1997  (17)  &14522 \\
11 & Wes  Unseld & 1980  (13)  &13769 & 11(16)  & 31 & Shaquille  O'neal & 2008  (17)  &14414 \\
12 & Hakeem  Olajuwon & 2001  (18)  &13747 & 12(18)  & 33 & Dikembe  Mutombo & 2008  (18)  &14148 \\
13 & Buck  Williams & 1997  (17)  &13018 & 13(17)  & 23 & Charles  Barkley & 1999  (16)  &14120 \\
14 & Jerry  Lucas & 1973  (11)  &12942 & 14(20)  & 30 & Charles  Oakley & 2003  (19)  &13740 \\
15 & Bob  Pettit & 1964  (11)  &12849 & 15(21)  & 28 & Dennis  Rodman & 1999  (14)  &13515 \\
16 & Shaquille  O'neal & 2008  (17)  &12566 & 16(23)  & 30 & Kevin  Garnett & 2008  (14)  &13479 \\
17 & Charles  Barkley & 1999  (16)  &12546 & 17(11)  & -54 & Wes  Unseld & 1980  (13)  &13439 \\
18 & Dikembe  Mutombo & 2008  (18)  &12359 & 18(22)  & 18 & Kevin  Willis & 2006  (21)  &13424 \\
19 & Paul  Silas & 1979  (16)  &12357 & 19(24)  & 20 & Patrick  Ewing & 2001  (17)  &13099 \\
20 & Charles  Oakley & 2003  (19)  &12205 & 20(9)  & -122 & Nate  Thurmond & 1976  (14)  &12891 \\
21 & Dennis  Rodman & 1999  (14)  &11954 & 21(10)  & -110 & Walt  Bellamy & 1974  (14)  &12219 \\
22 & Kevin  Willis & 2006  (21)  &11901 & 22(30)  & 26 & Tim  Duncan & 2008  (12)  &12151 \\
23 & Kevin  Garnett & 2008  (14)  &11682 & 23(32)  & 28 & David  Robinson & 2002  (14)  &11915 \\
24 & Patrick  Ewing & 2001  (17)  &11606 & 24(28)  & 14 & Jack  Sikma & 1990  (14)  &11839 \\
25 & Elgin  Baylor & 1971  (14)  &11463 & 25(35)  & 28 & Otis  Thorpe & 2000  (17)  &11667 \\
26 & Dan  Issel & 1984  (15)  &11133 & 26(19)  & -36 & Paul  Silas & 1979  (16)  &11657 \\
27 & Bill  Bridges & 1974  (13)  &11054 & 27(34)  & 20 & Bill  Laimbeer & 1993  (14)  &11513 \\
28 & Jack  Sikma & 1990  (14)  &10816 & 28(26)  & -7 & Dan  Issel & 1984  (15)  &11361 \\
29 & Caldwell  Jones & 1989  (17)  &10685 & 29(29)  & 0 & Caldwell  Jones & 1989  (17)  &11347 \\
30 & Tim  Duncan & 2008  (12)  &10546 & 30(14)  & -114 & Jerry  Lucas & 1973  (11)  &11112 \\
31 & Julius  Erving & 1986  (16)  &10525 & 31(31)  & 0 & Julius  Erving & 1986  (16)  &10873 \\
32 & David  Robinson & 2002  (14)  &10497 & 32(43)  & 25 & Horace  Grant & 2003  (17)  &10697 \\
33 & Dave  Cowens & 1982  (11)  &10444 & 33(42)  & 21 & A.c.  Green & 2000  (16)  &10685 \\
34 & Bill  Laimbeer & 1993  (14)  &10400 & 34(47)  & 27 & Ben  Wallace & 2008  (13)  &10643 \\
35 & Otis  Thorpe & 2000  (17)  &10370 & 35(45)  & 22 & Vlade  Divac & 2004  (16)  &10637 \\
36 & Johnny  Kerr & 1965  (12)  &10092 & 36(15)  & -140 & Bob  Pettit & 1964  (11)  &10436 \\
37 & Bob  Lanier & 1983  (14)  &9698 & 37(33)  & -12 & Dave  Cowens & 1982  (11)  &10351 \\
38 & Sam  Lacey & 1982  (13)  &9687 & 38(52)  & 26 & Shawn  Kemp & 2002  (14)  &10074 \\
39 & Zelmo  Beaty & 1974  (12)  &9665 & 39(46)  & 15 & Maurice  Lucas & 1987  (14)  &9945 \\
40 & Dave  Debusschere & 1973  (12)  &9618 & 40(50)  & 20 & Larry  Bird & 1991  (13)  &9908 \\
41 & Mel  Daniels & 1976  (9)  &9528 & 41(57)  & 28 & Dale  Davis & 2006  (16)  &9851 \\
42 & A.c.  Green & 2000  (16)  &9473 & 42(37)  & -13 & Bob  Lanier & 1983  (14)  &9790 \\
43 & Horace  Grant & 2003  (17)  &9443 & 43(38)  & -13 & Sam  Lacey & 1982  (13)  &9732 \\
44 & Bailey  Howell & 1970  (12)  &9383 & 44(55)  & 20 & Michael  Cage & 1999  (15)  &9698 \\
45 & Vlade  Divac & 2004  (16)  &9326 & 45(27)  & -66 & Bill  Bridges & 1974  (13)  &9682 \\
46 & Maurice  Lucas & 1987  (14)  &9306 & 46(59)  & 22 & P.j.  Brown & 2007  (15)  &9679 \\
47 & Ben  Wallace & 2008  (13)  &9243 & 47(56)  & 16 & Terry  Cummings & 1999  (18)  &9642 \\
48 & George  Mcginnis & 1981  (11)  &9233 & 48(48)  & 0 & George  Mcginnis & 1981  (11)  &9413 \\
49 & Johnny  Green & 1972  (14)  &9083 & 49(61)  & 19 & Chris  Webber & 2007  (15)  &9356 \\
50 & Larry  Bird & 1991  (13)  &8974 & 50(25)  & -100 & Elgin  Baylor & 1971  (14)  &9280 \\
\hline
\end{tabular}}}
\caption{  Ranking of Career Rebounds. 
{\scriptsize The left columns lists the traditional ranking of career
statistics, where the top 50 players are ranked along with their final season (career length in seasons listed in
parenthesis) and their career metric tally. The right columns list the renormalized ranking of career statistics
$Rank^{*}$, where the corresponding traditional ranking of the player is denoted in parenthesis. $L$ denotes the career
length of the player.  } }
\label{table:Creb}
\end{table}

\begin{table}[h]
\centering{ {\footnotesize
\begin{tabular}{@{\vrule height .5 pt depth4pt  width0pt}lccc||lcccc}
&\multicolumn4c{{\bf Traditional Rank}}&\multicolumn4c{{\bf Renormalized Rank}}\\
\noalign{
\vskip-1pt} Rank & Name & Final Season (L) & Career Metric & Rank$^{*}$(Rank)  &  \% Change & Name & Final Season (L)
& Career Metric \\
\hline 
1 & John  Stockton & 2002  (19)  &15806 & 1(1)  & 0 & John  Stockton & 2002  (19)  &15289 \\
2 & Mark  Jackson & 2003  (17)  &10323 & 2(3)  & 33 & Jason  Kidd & 2008  (15)  &10841 \\
3 & Jason  Kidd & 2008  (15)  &10199 & 3(2)  & -50 & Mark  Jackson & 2003  (17)  &10222 \\
4 & Magic  Johnson & 1995  (13)  &10141 & 4(5)  & 20 & Oscar  Robertson & 1973  (14)  &10144 \\
5 & Oscar  Robertson & 1973  (14)  &9887 & 5(7)  & 28 & Gary  Payton & 2006  (17)  &9229 \\
6 & Isiah  Thomas & 1993  (13)  &9061 & 6(4)  & -50 & Magic  Johnson & 1995  (13)  &9145 \\
7 & Gary  Payton & 2006  (17)  &8964 & 7(6)  & -16 & Isiah  Thomas & 1993  (13)  &8190 \\
8 & Rod  Strickland & 2004  (17)  &7987 & 8(9)  & 11 & Steve  Nash & 2008  (13)  &8090 \\
9 & Steve  Nash & 2008  (13)  &7504 & 9(8)  & -12 & Rod  Strickland & 2004  (17)  &8005 \\
10 & Maurice  Cheeks & 1992  (15)  &7392 & 10(11)  & 9 & Lenny  Wilkens & 1974  (15)  &7407 \\
11 & Lenny  Wilkens & 1974  (15)  &7211 & 11(14)  & 21 & Guy  Rodgers & 1969  (12)  &7183 \\
12 & Terry  Porter & 2001  (17)  &7160 & 12(13)  & 7 & Tim  Hardaway & 2002  (13)  &7064 \\
13 & Tim  Hardaway & 2002  (13)  &7095 & 13(19)  & 31 & Stephon  Marbury & 2008  (13)  &6920 \\
14 & Guy  Rodgers & 1969  (12)  &6917 & 14(12)  & -16 & Terry  Porter & 2001  (17)  &6800 \\
15 & Muggsy  Bogues & 2000  (14)  &6726 & 15(10)  & -50 & Maurice  Cheeks & 1992  (15)  &6653 \\
16 & Kevin  Johnson & 1999  (12)  &6711 & 16(27)  & 40 & Andre  Miller & 2008  (10)  &6469 \\
17 & Derek  Harper & 1998  (16)  &6571 & 17(15)  & -13 & Muggsy  Bogues & 2000  (14)  &6435 \\
18 & Nate  Archibald & 1983  (13)  &6476 & 18(16)  & -12 & Kevin  Johnson & 1999  (12)  &6390 \\
19 & Stephon  Marbury & 2008  (13)  &6471 & 19(23)  & 17 & Jerry  West & 1973  (14)  &6340 \\
20 & John  Lucas & 1989  (14)  &6454 & 20(29)  & 31 & Sam  Cassell & 2008  (16)  &6260 \\
21 & Reggie  Theus & 1990  (13)  &6453 & 21(26)  & 19 & John  Havlicek & 1977  (16)  &6164 \\
22 & Norm  Nixon & 1988  (10)  &6386 & 22(17)  & -29 & Derek  Harper & 1998  (16)  &6162 \\
23 & Jerry  West & 1973  (14)  &6238 & 23(18)  & -27 & Nate  Archibald & 1983  (13)  &6088 \\
24 & Scottie  Pippen & 2003  (17)  &6135 & 24(24)  & 0 & Scottie  Pippen & 2003  (17)  &6079 \\
25 & Clyde  Drexler & 1997  (15)  &6125 & 25(31)  & 19 & Nick  Vanexel & 2005  (13)  &6002 \\
26 & John  Havlicek & 1977  (16)  &6114 & 26(28)  & 7 & Mookie  Blaylock & 2001  (13)  &5934 \\
27 & Andre  Miller & 2008  (10)  &6020 & 27(35)  & 22 & Allen  Iverson & 2008  (13)  &5911 \\
28 & Mookie  Blaylock & 2001  (13)  &5972 & 28(30)  & 6 & Avery  Johnson & 2003  (16)  &5891 \\
29 & Sam  Cassell & 2008  (16)  &5939 & 29(20)  & -45 & John  Lucas & 1989  (14)  &5831 \\
30 & Avery  Johnson & 2003  (16)  &5846 & 30(21)  & -42 & Reggie  Theus & 1990  (13)  &5787 \\
31 & Nick  Vanexel & 2005  (13)  &5777 & 31(25)  & -24 & Clyde  Drexler & 1997  (15)  &5731 \\
32 & Larry  Bird & 1991  (13)  &5695 & 32(22)  & -45 & Norm  Nixon & 1988  (10)  &5719 \\
33 & Kareem  Abdul-jabbar & 1988  (20)  &5660 & 33(38)  & 13 & Damon  Stoudamire & 2007  (13)  &5689 \\
34 & Michael  Jordan & 2002  (15)  &5633 & 34(37)  & 8 & Dave  Bing & 1977  (12)  &5428 \\
35 & Allen  Iverson & 2008  (13)  &5512 & 35(34)  & -2 & Michael  Jordan & 2002  (15)  &5359 \\
36 & Dennis  Johnson & 1989  (14)  &5499 & 36(33)  & -9 & Kareem  Abdul-jabbar & 1988  (20)  &5290 \\
37 & Dave  Bing & 1977  (12)  &5397 & 37(49)  & 24 & Baron  Davis & 2008  (10)  &5266 \\
38 & Damon  Stoudamire & 2007  (13)  &5371 & 38(44)  & 13 & Kenny  Anderson & 2004  (14)  &5252 \\
39 & Kevin  Porter & 1982  (10)  &5314 & 39(53)  & 26 & Mike  Bibby & 2008  (11)  &5226 \\
40 & Jeff  Hornacek & 1999  (14)  &5281 & 40(41)  & 2 & Karl  Malone & 2003  (19)  &5220 \\
41 & Karl  Malone & 2003  (19)  &5248 & 41(32)  & -28 & Larry  Bird & 1991  (13)  &5132 \\
42 & Rickey  Green & 1991  (14)  &5221 & 42(40)  & -5 & Jeff  Hornacek & 1999  (14)  &5064 \\
43 & Norm  Vanlier & 1978  (10)  &5217 & 43(47)  & 8 & Walt  Frazier & 1979  (13)  &5050 \\
44 & Kenny  Anderson & 2004  (14)  &5196 & 44(43)  & -2 & Norm  Vanlier & 1978  (10)  &5049 \\
45 & Julius  Erving & 1986  (16)  &5176 & 45(58)  & 22 & Chauncey  Billups & 2008  (12)  &5048 \\
46 & Sleepy  Floyd & 1994  (13)  &5175 & 46(36)  & -27 & Dennis  Johnson & 1989  (14)  &4940 \\
47 & Walt  Frazier & 1979  (13)  &5040 & 47(39)  & -20 & Kevin  Porter & 1982  (10)  &4928 \\
48 & Rick  Barry & 1979  (14)  &4952 & 48(59)  & 18 & Wilt  Chamberlain & 1972  (14)  &4848 \\
49 & Baron  Davis & 2008  (10)  &4902 & 49(64)  & 23 & Kevin  Garnett & 2008  (14)  &4847 \\
50 & Nate  Mcmillan & 1997  (12)  &4893 & 50(66)  & 24 & Brevin  Knight & 2008  (12)  &4815 \\
\hline
\end{tabular}}}
\caption{  Ranking of Career Assists. 
{\scriptsize The left columns lists the traditional ranking of career
statistics, where the top 50 players are ranked along with their final season (career length in seasons listed in
parenthesis) and their career metric tally. The right columns list the renormalized ranking of career statistics
$Rank^{*}$, where the corresponding traditional ranking of the player is denoted in parenthesis. $L$ denotes the career
length of the player.  } }
\label{table:Cast}
\end{table}

\begin{table}[h]
\centering{ {\footnotesize
\begin{tabular}{@{\vrule height .5 pt depth4pt  width0pt}lccc||lcccc}
&\multicolumn4c{{\bf Traditional Rank}}&\multicolumn4c{{\bf Renormalized Rank}}\\
\noalign{
\vskip-1pt} Rank & Name & Season ($Y\#$) & Season Metric & Rank$^{*}$(Rank)  &  \% Change & Name & Season ($Y\#$) &
Season Metric \\
\hline 
1 & Wilt  Chamberlain & 1961  (3)  &4029 & 1(1)  & 0 & Wilt  Chamberlain & 1961  (3)  &3543 \\
2 & Wilt  Chamberlain & 1962  (4)  &3586 & 2(2)  & 0 & Wilt  Chamberlain & 1962  (4)  &3248 \\
3 & Michael  Jordan & 1986  (3)  &3041 & 3(7)  & 57 & Kobe  Bryant & 2005  (10)  &3060 \\
4 & Wilt  Chamberlain & 1960  (2)  &3033 & 4(3)  & -33 & Michael  Jordan & 1986  (3)  &2892 \\
5 & Wilt  Chamberlain & 1963  (5)  &2948 & 5(8)  & 37 & Bob  Mcadoo & 1974  (3)  &2823 \\
6 & Michael  Jordan & 1987  (4)  &2868 & 6(5)  & -20 & Wilt  Chamberlain & 1963  (5)  &2771 \\
7 & Kobe  Bryant & 2005  (10)  &2832 & 7(6)  & -16 & Michael  Jordan & 1987  (4)  &2769 \\
8 & Bob  Mcadoo & 1974  (3)  &2831 & 8(37)  & 78 & Kobe  Bryant & 2002  (7)  &2711 \\
9 & Kareem  Abdul-jabbar & 1971  (3)  &2822 & 9(11)  & 18 & Michael  Jordan & 1989  (6)  &2690 \\
10 & Rick  Barry & 1966  (2)  &2775 & 10(4)  & -150 & Wilt  Chamberlain & 1960  (2)  &2681 \\
11 & Michael  Jordan & 1989  (6)  &2753 & 11(34)  & 67 & LeBron  James & 2005  (3)  &2677 \\
12 & Elgin  Baylor & 1962  (5)  &2719 & 12(49)  & 75 & Tracy  Mcgrady & 2002  (6)  &2651 \\
12 & Nate  Archibald & 1972  (3)  &2719 & 13(9)  & -44 & Kareem  Abdul-jabbar & 1971  (3)  &2646 \\
14 & Wilt  Chamberlain & 1959  (1)  &2707 & 14(55)  & 74 & Jerry  Stackhouse & 2000  (6)  &2629 \\
15 & Wilt  Chamberlain & 1965  (7)  &2649 & 15(42)  & 64 & Michael  Jordan & 1996  (12)  &2625 \\
16 & Charlie  Scott & 1971  (2)  &2637 & 16(31)  & 48 & Michael  Jordan & 1995  (11)  &2618 \\
17 & Michael  Jordan & 1988  (5)  &2633 & 17(12)  & -41 & Nate  Archibald & 1972  (3)  &2598 \\
18 & Kareem  Abdul-jabbar & 1970  (2)  &2596 & 18(62)  & 70 & Michael  Jordan & 1997  (13)  &2582 \\
19 & George  Gervin & 1979  (8)  &2585 & 19(43)  & 55 & Kobe  Bryant & 2006  (11)  &2580 \\
20 & Michael  Jordan & 1990  (7)  &2580 & 20(57)  & 64 & Allen  Iverson & 2005  (10)  &2568 \\
21 & George  Gervin & 1981  (10)  &2551 & 21(20)  & -5 & Michael  Jordan & 1990  (7)  &2540 \\
22 & Michael  Jordan & 1992  (9)  &2541 & 22(66)  & 66 & Gilbert  Arenas & 2005  (5)  &2535 \\
23 & Karl  Malone & 1989  (5)  &2540 & 23(22)  & -4 & Michael  Jordan & 1992  (9)  &2525 \\
24 & Dan  Issel & 1971  (2)  &2538 & 24(17)  & -41 & Michael  Jordan & 1988  (5)  &2522 \\
24 & Elgin  Baylor & 1960  (3)  &2538 & 25(68)  & 63 & Shaquille  O'neal & 1999  (8)  &2514 \\
26 & Wilt  Chamberlain & 1964  (6)  &2534 & 26(52)  & 50 & Dwyane  Wade & 2008  (6)  &2498 \\
27 & Moses  Malone & 1981  (8)  &2520 & 27(105)  & 74 & Allen  Iverson & 2002  (7)  &2492 \\
28 & Spencer  Haywood & 1969  (1)  &2519 & 28(23)  & -21 & Karl  Malone & 1989  (5)  &2482 \\
29 & Rick  Barry & 1971  (6)  &2518 & 29(84)  & 65 & Allen  Iverson & 2004  (9)  &2479 \\
30 & Walt  Bellamy & 1961  (1)  &2495 & 30(19)  & -57 & George  Gervin & 1979  (8)  &2475 \\
31 & Michael  Jordan & 1995  (11)  &2491 & 31(16)  & -93 & Charlie  Scott & 1971  (2)  &2472 \\
32 & Oscar  Robertson & 1963  (4)  &2480 & 32(10)  & -220 & Rick  Barry & 1966  (2)  &2468 \\
32 & Dan  Issel & 1970  (1)  &2480 & 33(12)  & -175 & Elgin  Baylor & 1962  (5)  &2463 \\
34 & LeBron  James & 2005  (3)  &2478 & 34(21)  & -61 & George  Gervin & 1981  (10)  &2457 \\
35 & Jerry  West & 1965  (6)  &2476 & 35(53)  & 33 & David  Robinson & 1993  (5)  &2450 \\
36 & Julius  Erving & 1975  (5)  &2462 & 36(14)  & -157 & Wilt  Chamberlain & 1959  (1)  &2448 \\
37 & Kobe  Bryant & 2002  (7)  &2461 & 37(57)  & 35 & Shaquille  O'neal & 1993  (2)  &2444 \\
38 & Adrian  Dantley & 1981  (6)  &2457 & 38(40)  & 5 & Rick  Barry & 1974  (9)  &2443 \\
39 & Adrian  Dantley & 1980  (5)  &2452 & 39(126)  & 69 & Allen  Iverson & 2000  (5)  &2438 \\
40 & Rick  Barry & 1974  (9)  &2450 & 40(76)  & 47 & Kobe  Bryant & 2007  (12)  &2431 \\
41 & Oscar  Robertson & 1961  (2)  &2432 & 41(115)  & 64 & Karl  Malone & 1996  (12)  &2428 \\
42 & Michael  Jordan & 1996  (12)  &2431 & 42(27)  & -55 & Moses  Malone & 1981  (8)  &2427 \\
43 & Kobe  Bryant & 2006  (11)  &2430 & 43(36)  & -19 & Julius  Erving & 1975  (5)  &2414 \\
44 & Bob  Pettit & 1961  (8)  &2429 & 44(82)  & 46 & LeBron  James & 2008  (6)  &2412 \\
45 & Bob  Mcadoo & 1975  (4)  &2427 & 45(131)  & 65 & Karl  Malone & 1997  (13)  &2399 \\
46 & Adrian  Dantley & 1983  (8)  &2418 & 46(26)  & -76 & Wilt  Chamberlain & 1964  (6)  &2396 \\
47 & Alex  English & 1985  (10)  &2414 & 47(77)  & 38 & Shaquille  O'neal & 1994  (3)  &2390 \\
48 & Oscar  Robertson & 1966  (7)  &2412 & 48(50)  & 4 & Michael  Jordan & 1991  (8)  &2389 \\
49 & Tracy  Mcgrady & 2002  (6)  &2407 & 49(15)  & -226 & Wilt  Chamberlain & 1965  (7)  &2388 \\
50 & Michael  Jordan & 1991  (8)  &2404 & 50(45)  & -11 & Bob  Mcadoo & 1975  (4)  &2379 \\
\hline
\end{tabular}}}
\caption{ Ranking of Season Points. 
The left columns list the traditional ranking
of season statistics, where the top 50 players are ranked along with the year. The right columns list the renormalized
ranking of season statistics $Rank^{*}$. $Y\#$ denotes the number of years into the career. }
\label{table:Spts}
\end{table}

\begin{table}[h]
\centering{ {\footnotesize
\begin{tabular}{@{\vrule height .5 pt depth4pt  width0pt}lccc||lcccc}
&\multicolumn4c{{\bf Traditional Rank}}&\multicolumn4c{{\bf Renormalized Rank}}\\
\noalign{
\vskip-1pt} Rank & Name & Season ($Y\#$) & Season Metric & Rank$^{*}$(Rank)  &  \% Change & Name & Season ($Y\#$) &
Season Metric \\
\hline 
1 & Wilt  Chamberlain & 1960  (2)  &2149 & 1(28)  & 96 & Dennis  Rodman & 1991  (6)  &1691 \\
2 & Wilt  Chamberlain & 1961  (3)  &2052 & 2(4)  & 50 & Wilt  Chamberlain & 1967  (9)  &1666 \\
3 & Wilt  Chamberlain & 1966  (8)  &1957 & 3(5)  & 40 & Wilt  Chamberlain & 1962  (4)  &1628 \\
4 & Wilt  Chamberlain & 1967  (9)  &1952 & 4(1)  & -300 & Wilt  Chamberlain & 1960  (2)  &1605 \\
5 & Wilt  Chamberlain & 1962  (4)  &1946 & 5(2)  & -150 & Wilt  Chamberlain & 1961  (3)  &1600 \\
6 & Wilt  Chamberlain & 1965  (7)  &1943 & 6(8)  & 25 & Bill  Russell & 1963  (8)  &1595 \\
7 & Wilt  Chamberlain & 1959  (1)  &1941 & 7(3)  & -133 & Wilt  Chamberlain & 1966  (8)  &1587 \\
8 & Bill  Russell & 1963  (8)  &1930 & 8(6)  & -33 & Wilt  Chamberlain & 1965  (7)  &1552 \\
9 & Bill  Russell & 1964  (9)  &1878 & 9(11)  & 18 & Bill  Russell & 1962  (7)  &1542 \\
10 & Bill  Russell & 1960  (5)  &1868 & 10(43)  & 76 & Moses  Malone & 1978  (5)  &1537 \\
11 & Bill  Russell & 1962  (7)  &1843 & 11(27)  & 59 & Artis  Gilmore & 1973  (3)  &1532 \\
12 & Bill  Russell & 1961  (6)  &1790 & 11(9)  & -22 & Bill  Russell & 1964  (9)  &1532 \\
13 & Wilt  Chamberlain & 1963  (5)  &1787 & 13(52)  & 75 & Dennis  Rodman & 1993  (8)  &1530 \\
14 & Bill  Russell & 1965  (10)  &1779 & 14(16)  & 12 & Wilt  Chamberlain & 1968  (10)  &1479 \\
15 & Bill  Russell & 1959  (4)  &1778 & 15(20)  & 25 & Spencer  Haywood & 1969  (1)  &1478 \\
16 & Wilt  Chamberlain & 1968  (10)  &1712 & 16(13)  & -23 & Wilt  Chamberlain & 1963  (5)  &1477 \\
17 & Bill  Russell & 1966  (11)  &1700 & 17(29)  & 41 & Wilt  Chamberlain & 1972  (14)  &1468 \\
18 & Wilt  Chamberlain & 1964  (6)  &1673 & 18(22)  & 18 & Wilt  Chamberlain & 1971  (13)  &1467 \\
19 & Jerry  Lucas & 1965  (3)  &1668 & 19(37)  & 48 & Elvin  Hayes & 1973  (6)  &1458 \\
20 & Spencer  Haywood & 1969  (1)  &1637 & 20(7)  & -185 & Wilt  Chamberlain & 1959  (1)  &1456 \\
21 & Bill  Russell & 1958  (3)  &1612 & 21(14)  & -50 & Bill  Russell & 1965  (10)  &1421 \\
22 & Wilt  Chamberlain & 1971  (13)  &1572 & 22(35)  & 37 & Artis  Gilmore & 1972  (2)  &1420 \\
23 & Bill  Russell & 1957  (2)  &1564 & 23(12)  & -91 & Bill  Russell & 1961  (6)  &1396 \\
24 & Jerry  Lucas & 1967  (5)  &1560 & 24(96)  & 75 & Dennis  Rodman & 1997  (12)  &1395 \\
25 & Jerry  Lucas & 1966  (4)  &1547 & 24(10)  & -140 & Bill  Russell & 1960  (5)  &1395 \\
26 & Bob  Pettit & 1960  (7)  &1540 & 26(32)  & 18 & Artis  Gilmore & 1971  (1)  &1391 \\
27 & Artis  Gilmore & 1973  (3)  &1538 & 27(73)  & 63 & Kevin  Willis & 1991  (7)  &1390 \\
28 & Dennis  Rodman & 1991  (6)  &1530 & 28(54)  & 48 & Artis  Gilmore & 1974  (4)  &1385 \\
29 & Wilt  Chamberlain & 1972  (14)  &1526 & 29(48)  & 39 & Kareem  Abdul-jabbar & 1975  (7)  &1384 \\
30 & Walt  Bellamy & 1961  (1)  &1500 & 30(17)  & -76 & Bill  Russell & 1966  (11)  &1379 \\
31 & Wilt  Chamberlain & 1970  (12)  &1493 & 31(18)  & -72 & Wilt  Chamberlain & 1964  (6)  &1365 \\
32 & Wes  Unseld & 1968  (1)  &1491 & 32(31)  & -3 & Wilt  Chamberlain & 1970  (12)  &1341 \\
32 & Artis  Gilmore & 1971  (1)  &1491 & 33(15)  & -120 & Bill  Russell & 1959  (4)  &1333 \\
34 & Bill  Russell & 1968  (13)  &1484 & 33(119)  & 72 & Dwight  Howard & 2007  (4)  &1333 \\
35 & Artis  Gilmore & 1972  (2)  &1476 & 35(19)  & -84 & Jerry  Lucas & 1965  (3)  &1332 \\
36 & Mel  Daniels & 1970  (4)  &1475 & 35(24)  & -45 & Jerry  Lucas & 1967  (5)  &1332 \\
37 & Elvin  Hayes & 1973  (6)  &1463 & 37(36)  & -2 & Mel  Daniels & 1970  (4)  &1325 \\
38 & Mel  Daniels & 1969  (3)  &1462 & 38(38)  & 0 & Mel  Daniels & 1969  (3)  &1320 \\
39 & Bob  Pettit & 1961  (8)  &1459 & 39(69)  & 43 & Truck  Robinson & 1977  (4)  &1319 \\
40 & Julius  Keye & 1970  (2)  &1454 & 40(106)  & 62 & Moses  Malone & 1981  (8)  &1318 \\
41 & Bill  Russell & 1967  (12)  &1451 & 41(111)  & 63 & Moses  Malone & 1980  (7)  &1307 \\
42 & Elgin  Baylor & 1960  (3)  &1447 & 42(87)  & 51 & Swen  Nater & 1979  (7)  &1306 \\
43 & Moses  Malone & 1978  (5)  &1444 & 42(40)  & -5 & Julius  Keye & 1970  (2)  &1306 \\
44 & Elvin  Hayes & 1968  (1)  &1406 & 44(67)  & 34 & Artis  Gilmore & 1975  (5)  &1304 \\
45 & Nate  Thurmond & 1968  (6)  &1402 & 45(128)  & 64 & Kevin  Garnett & 2003  (9)  &1302 \\
46 & Nate  Thurmond & 1964  (2)  &1395 & 45(70)  & 35 & Swen  Nater & 1974  (2)  &1302 \\
47 & Elvin  Hayes & 1969  (2)  &1386 & 47(120)  & 60 & Dikembe  Mutombo & 1999  (9)  &1299 \\
48 & Kareem  Abdul-jabbar & 1975  (7)  &1383 & 48(57)  & 15 & Nate  Thurmond & 1972  (10)  &1297 \\
49 & Nate  Thurmond & 1966  (4)  &1382 & 49(99)  & 50 & Moses  Malone & 1982  (9)  &1293 \\
50 & Jerry  Lucas & 1963  (1)  &1375 & 50(21)  & -138 & Bill  Russell & 1958  (3)  &1290 \\
\hline
\end{tabular}}}
\caption{ Ranking of Season Rebounds. 
 The left columns list the traditional ranking
of season statistics, where the top 50 players are ranked along with the year. The right columns list the renormalized
ranking of season statistics $Rank^{*}$. $Y\#$ denotes the number of years into the career. }
\label{table:Sreb}
\end{table}

\begin{table}[h]
\centering{ { \footnotesize
\begin{tabular}{@{\vrule height .5 pt depth4pt  width0pt}lccc||lcccc}
&\multicolumn4c{{\bf Traditional Rank}}&\multicolumn4c{{\bf Renormalized Rank}}\\
\noalign{
\vskip-1pt} Rank & Name & Season ($Y\#$) & Season Metric & Rank$^{*}$(Rank)  &  \% Change & Name & Season ($Y\#$) &
Season Metric \\
\hline 
1 & John  Stockton & 1990  (7)  &1164 & 1(1)  & 0 & John  Stockton & 1990  (7)  &1085 \\
2 & John  Stockton & 1989  (6)  &1134 & 2(4)  & 50 & John  Stockton & 1991  (8)  &1061 \\
3 & John  Stockton & 1987  (4)  &1128 & 3(2)  & -50 & John  Stockton & 1989  (6)  &1052 \\
4 & John  Stockton & 1991  (8)  &1126 & 4(3)  & -33 & John  Stockton & 1987  (4)  &1007 \\
5 & Isiah  Thomas & 1984  (4)  &1123 & 4(6)  & 33 & John  Stockton & 1988  (5)  &1007 \\
6 & John  Stockton & 1988  (5)  &1118 & 6(9)  & 33 & John  Stockton & 1994  (11)  &998 \\
7 & Kevin  Porter & 1978  (7)  &1099 & 7(5)  & -40 & Isiah  Thomas & 1984  (4)  &985 \\
8 & John  Stockton & 1993  (10)  &1031 & 8(7)  & -14 & Kevin  Porter & 1978  (7)  &982 \\
9 & John  Stockton & 1994  (11)  &1011 & 9(16)  & 43 & Mark  Jackson & 1996  (10)  &979 \\
10 & Kevin  Johnson & 1988  (2)  &991 & 9(17)  & 47 & Chris  Paul & 2007  (3)  &979 \\
11 & Magic  Johnson & 1990  (12)  &989 & 11(8)  & -37 & John  Stockton & 1993  (10)  &972 \\
12 & Magic  Johnson & 1988  (10)  &988 & 12(27)  & 55 & Steve  Nash & 2006  (11)  &959 \\
13 & John  Stockton & 1992  (9)  &987 & 13(26)  & 50 & Steve  Nash & 2007  (12)  &949 \\
14 & Magic  Johnson & 1986  (8)  &977 & 14(34)  & 58 & Oscar  Robertson & 1964  (5)  &947 \\
15 & Magic  Johnson & 1984  (6)  &968 & 15(34)  & 55 & Chris  Paul & 2008  (4)  &945 \\
16 & Mark  Jackson & 1996  (10)  &935 & 16(34)  & 52 & Steve  Nash & 2004  (9)  &933 \\
17 & Chris  Paul & 2007  (3)  &925 & 17(30)  & 43 & Oscar  Robertson & 1963  (4)  &932 \\
18 & John  Stockton & 1995  (12)  &916 & 17(22)  & 22 & Guy  Rodgers & 1966  (9)  &932 \\
19 & Isiah  Thomas & 1983  (3)  &914 & 19(18)  & -5 & John  Stockton & 1995  (12)  &930 \\
19 & Norm  Nixon & 1983  (7)  &914 & 20(28)  & 28 & Andre  Miller & 2001  (3)  &928 \\
21 & Nate  Archibald & 1972  (3)  &910 & 21(51)  & 58 & Steve  Nash & 2005  (10)  &925 \\
22 & Guy  Rodgers & 1966  (9)  &908 & 22(11)  & -100 & Magic  Johnson & 1990  (12)  &922 \\
23 & Magic  Johnson & 1989  (11)  &907 & 23(13)  & -76 & John  Stockton & 1992  (9)  &921 \\
23 & Magic  Johnson & 1985  (7)  &907 & 24(33)  & 27 & Deron  Williams & 2007  (3)  &912 \\
25 & Oscar  Robertson & 1961  (2)  &899 & 25(37)  & 32 & John  Stockton & 1996  (13)  &901 \\
26 & Steve  Nash & 2007  (12)  &897 & 26(10)  & -160 & Kevin  Johnson & 1988  (2)  &893 \\
27 & Steve  Nash & 2006  (11)  &884 & 27(12)  & -125 & Magic  Johnson & 1988  (10)  &890 \\
28 & Andre  Miller & 2001  (3)  &882 & 28(43)  & 34 & Oscar  Robertson & 1966  (7)  &867 \\
29 & Magic  Johnson & 1983  (5)  &875 & 29(14)  & -107 & Magic  Johnson & 1986  (8)  &866 \\
30 & Oscar  Robertson & 1963  (4)  &868 & 30(25)  & -20 & Oscar  Robertson & 1961  (2)  &865 \\
30 & Mark  Jackson & 1987  (1)  &868 & 31(58)  & 46 & Jason  Kidd & 2007  (14)  &853 \\
32 & Muggsy  Bogues & 1989  (3)  &867 & 32(56)  & 42 & Jason  Kidd & 2001  (8)  &850 \\
33 & Deron  Williams & 2007  (3)  &862 & 32(40)  & 20 & Oscar  Robertson & 1965  (6)  &850 \\
34 & Oscar  Robertson & 1964  (5)  &861 & 32(21)  & -52 & Nate  Archibald & 1972  (3)  &850 \\
34 & Chris  Paul & 2008  (4)  &861 & 35(15)  & -133 & Magic  Johnson & 1984  (6)  &849 \\
34 & Steve  Nash & 2004  (9)  &861 & 35(41)  & 14 & Guy  Rodgers & 1965  (8)  &849 \\
37 & John  Stockton & 1996  (13)  &860 & 37(68)  & 45 & Oscar  Robertson & 1968  (9)  &844 \\
38 & Magic  Johnson & 1987  (9)  &858 & 38(23)  & -65 & Magic  Johnson & 1989  (11)  &841 \\
39 & Sleepy  Floyd & 1986  (5)  &848 & 39(59)  & 33 & Rod  Strickland & 1997  (10)  &839 \\
40 & Oscar  Robertson & 1965  (6)  &847 & 40(52)  & 23 & Guy  Rodgers & 1962  (5)  &836 \\
41 & Guy  Rodgers & 1965  (8)  &846 & 41(114)  & 64 & Wilt  Chamberlain & 1967  (9)  &835 \\
41 & Kevin  Johnson & 1989  (3)  &846 & 42(46)  & 8 & Norm  Vanlier & 1970  (2)  &816 \\
43 & Oscar  Robertson & 1966  (7)  &845 & 43(83)  & 48 & Deron  Williams & 2006  (2)  &808 \\
44 & Kevin  Porter & 1977  (6)  &837 & 43(141)  & 69 & Lenny  Wilkens & 1967  (8)  &808 \\
45 & Kevin  Johnson & 1991  (5)  &836 & 45(19)  & -136 & Isiah  Thomas & 1983  (3)  &806 \\
46 & Norm  Vanlier & 1970  (2)  &832 & 45(19)  & -136 & Norm  Nixon & 1983  (7)  &806 \\
46 & Michealray  Richardson & 1979  (2)  &832 & 47(23)  & -104 & Magic  Johnson & 1985  (7)  &804 \\
48 & Terry  Porter & 1987  (3)  &831 & 47(32)  & -46 & Muggsy  Bogues & 1989  (3)  &804 \\
49 & Isiah  Thomas & 1985  (5)  &830 & 49(61)  & 19 & Avery  Johnson & 1995  (8)  &801 \\
50 & Magic  Johnson & 1982  (4)  &829 & 50(90)  & 44 & Jason  Kidd & 2006  (13)  &799 \\
\hline
\end{tabular}}}
\caption{ Ranking of Season Assists.  
The left columns list the traditional ranking
of season statistics, where the top 50 players are ranked along with the year. The right columns list the renormalized
ranking of season statistics $Rank^{*}$. $Y\#$ denotes the number of years into the career. }
\label{table:Sast}
\end{table}

\end{document}